\pdfoutput=1

\newif\ifdraft\drafttrue
\newif\ifcolor\colortrue

\documentclass{usiinftr}

\usepackage{lastpage}

\usepackage{xcolor}
\definecolor{pennblue}{cmyk}{1,.65,0,.3}
\definecolor{pennred}{cmyk}{0,1,.65,.34}

\usepackage{cmtt}

\usepackage{paralist}
\usepackage{lastpage}
\usepackage{times}
\usepackage{tikz}
\usepackage{alltt}
\usepackage{graphicx}
\usepackage{balance}
\usepackage{amsmath}
\usepackage{caption}
\usepackage{subcaption}
\usepackage{xspace}

\graphicspath{{./figures/}{../figures/}}

\usepackage[linesnumbered,ruled,vlined]{algorithm2e}
\SetAlFnt{\small}

\usepackage{url}

\usepackage{listings}

\usepackage{booktabs}
\newcommand{\ra}[1]{\renewcommand{\arraystretch}{#1}}

\begin{document}

\title{Optimistic Aborts for Geo-distributed Transactions}

\author{Theo Jepsen}{}
\author{Leandro Pacheco de Sousa}{}
\author{Huynh Tu Dang}{}
\author{\\Fernando Pedone}{} 
\author{Robert Soul\'{e}}{}

\affiliation{}{\USIINF}

\TRnumber{2016-05}

\date{}

\maketitle

\begin{abstract}
Network latency can have a significant impact on the performance of
transactional storage systems, particularly in wide area or geo-distributed
deployments. To reduce latency, systems typically rely on a cache to service
read-requests closer to the client. However, caches are not effective for
write-heavy workloads, which have to be processed by the storage system in order
to maintain serializability.

This paper presents a new technique, called \emph{optimistic abort}, which
reduces network latency for high-contention, write-heavy workloads by
identifying transactions that will abort as early as possible, and aborting them
before they reach the store. We have implemented \emph{optimistic abort} in a
system called Gotthard, which leverages recent advances in network data plane
programmability to execute transaction processing logic directly in network
devices. Gotthard examines network traffic to observe and log transaction
requests. If Gotthard suspects that a transaction is likely to be aborted at the
store, it aborts the transaction early by re-writing the packet header, and
routing the packets back to the client. Gotthard significantly reduces the
overall latency and improves the throughput for high-contention workloads.

\end{abstract}

\section{Introduction}
\label{sec:introduction}

Many distributed applications produce write-intensive workloads that access
geographically-distributed storage systems. Examples of such applications span a
range of domains, from popular web services offered by companies such as Google,
Microsoft, and Amazon~\cite{campbell2010extreme, baker11megastore, corbett12, calder2011windows, decandia07},
to sensor-network based monitoring for military purposes~\cite{durisic12} or
environmental tracking~\cite{cardell-oliver05,tolle05,werner-allen06}.

For these applications, network latency can have a significant impact on 
performance. The standard approach for reducing network latency is to deploy
proxy servers that service read operations closer to the client with cached
data~\cite{freedman04,akamai,nishtala2013scaling}.  Unfortunately, although
caching works well for read-heavy workloads~\cite{li16}, it provides little
benefit for write-heavy workloads, since write operations must be routed
directly to the store to ensure serializability
(\S\ref{sec:motivation}).

This paper presents a new technique, called \emph{optimistic abort}, which
reduces network latency for high-contention, write-heavy workloads. The key idea
is to identify transactions that will abort as early as possible, and abort them
before they reach the store. Optimistic abort provides a complimentary technique
to caching; rather than moving data closer to the client, it \emph{aborts
transactions closer to the client}.

The logic of optimistic abort could be realized in various ways. For example,
similar to a read cache, a proxy server could be used to identify and abort
transactions. In this paper, we explore a highly efficient, switch-based
implementation that leverages recent advances in data plane
programmability~\cite{song13,brebner14,bosshart14}.  Specifically, our system,
is implemented in the P4 language~\cite{bosshart14}, allowing it to use the
emerging ecosystem of compilers and tools to run on reconfigurable network
hardware~\cite{wang15,open-nfp,sdnet,p4c}.

With the resulting system, named Gotthard, network devices naturally serve as
message aggregation points for multiple clients.  Clients issue transaction
requests using a custom network protocol header. Gotthard switches examine
network traffic to observe and log transaction requests. If Gotthard suspects
that the store is likely to abort a transaction, the Gotthard switch proactively
aborts the transaction by re-writing the packet header, and forwarding the
packet back to the client. Since Gotthard aborts the transaction before the
packet travels the network to the storage system, it can significantly reduce
the overall latency for processing the request. Moreover, Gotthard's optimistic
strategy reduces commit time, improving overall system throughput.

We have implemented a prototype of Gotthard, and evaluated it both in emulation
using Mininet~\cite{mininet}, and on Amazon's EC2 cloud infrastructure. Our
experiments include both a set of microbenchmarks that explore the parameter
space for different operational conditions, and an implementation of
TPC-C~\cite{tpcc} to emulate a real-world inspired workload.  Our evaluation
shows that Gotthard significantly reduces the latency for transactions, and
under high-contention workloads, and increases transaction throughput.

As a highlight of some of results, our evaluation shows that when
deployed on regions in the western and eastern United States in EC2,
Gotthard almost doubles the throughput for TPC-C Payment transactions.

Overall, this paper makes the following contributions:
\begin{itemize}
\item It presents a novel algorithm for improving the performance of
transaction systems by optimistically aborting transactions before they reach the store. 
\item It describes an implementation of the optimistic abort technique 
that builds on emerging technological trends in data plane programming languages.
\item It explores the parameter space for network-based
  transaction processing, and demonstrates significant
performance improvements for realistic workloads.
\end{itemize}

\noindent
The rest of this paper is organized as follows.  We first motivate Gotthard with
an example experiment (\S\ref{sec:motivation}). We then present the design of
the Gotthard system in detail (\S\ref{sec:design}). Next, we describe the
implementation (\S\ref{sec:implementation}) and present a thorough evaluation
(\S\ref{sec:evaluation}).  Finally, we discuss related work
(\S\ref{sec:related}), and conclude (\S\ref{sec:conclusion}).

\section{Background and Motivation}
\label{sec:motivation}

Before detailing the design of Gotthard, we briefly describe opportunities due
to technological trends, the system model, and present an experiment that
motivates moving transaction processing logic into network devices.

\subsection{Opportunity}

Recently, the landscape for network computing hardware has changed dramatically.
 Several devices are on the horizon that offer flexible hardware
with customizable packet processing pipelines, including Protocol Independent
Switch Architecture (PISA) chips from Barefoot networks~\cite{bosshart13},
FlexPipe from Intel~\cite{jose15}, NFP-6xxx from
Netronome~\cite{netronome-hotchips}, and Xpliant from Cavium~\cite{xpliant}.

Importantly, this hardware is not simply more powerful; it is also more
programmable. Several vendors and consortia have proposed domain specific
languages that target these devices~\cite{song13,brebner14,bosshart14}. These
languages allow users to customize their network hardware with new applications
and services. Particularly relevant to Gotthard is the P4
language~\cite{bosshart14}. P4 provides high-level abstractions that are
tailored directly to the needs of network forwarding devices: packets are
processed by a sequence of tables; tables match header fields, and perform
actions that forward, drop, or modify packets.  Moreover, P4 allows for stateful
operations that can read and write to memory cells called registers.

As a result of these changes, networks now offer a programmable substrate that
was not previously available to system designers. Consequently, there is an
opportunity to re-visit the traditional separation of application-layer
and transport layer logic advocated by the end-to-end principle~\cite{saltzer84}.
In the case of Gotthard, we leverage the programmable network substrate to
improve performance for write-heavy workloads.

\subsection{System Model}

\begin{figure}[t]
    \centering
    \begin{minipage}{.45\textwidth}
        \centering
         \includegraphics[width=\textwidth]{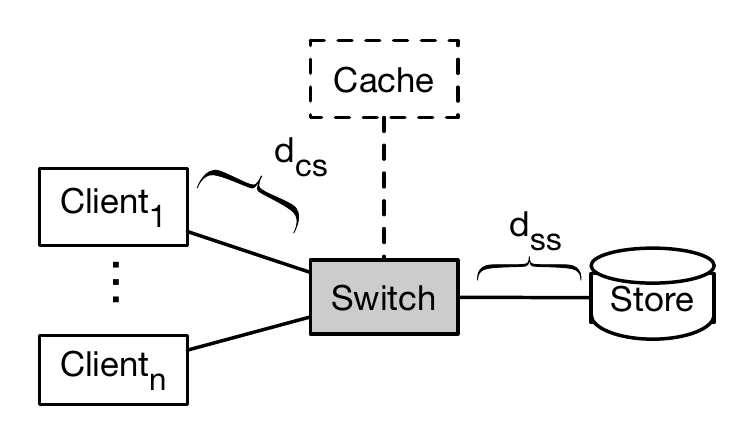}
        \caption{A typical cache would run on a server connected to a switch. For our
          experiments, we have implemented an idealized cache deployment that runs
          directly in the switch itself, thus reducing network latency.}
        \label{fig:cache}
    \end{minipage}%
    \hfill
    \begin{minipage}{.45\textwidth}
        \centering
         \includegraphics[width=\textwidth]{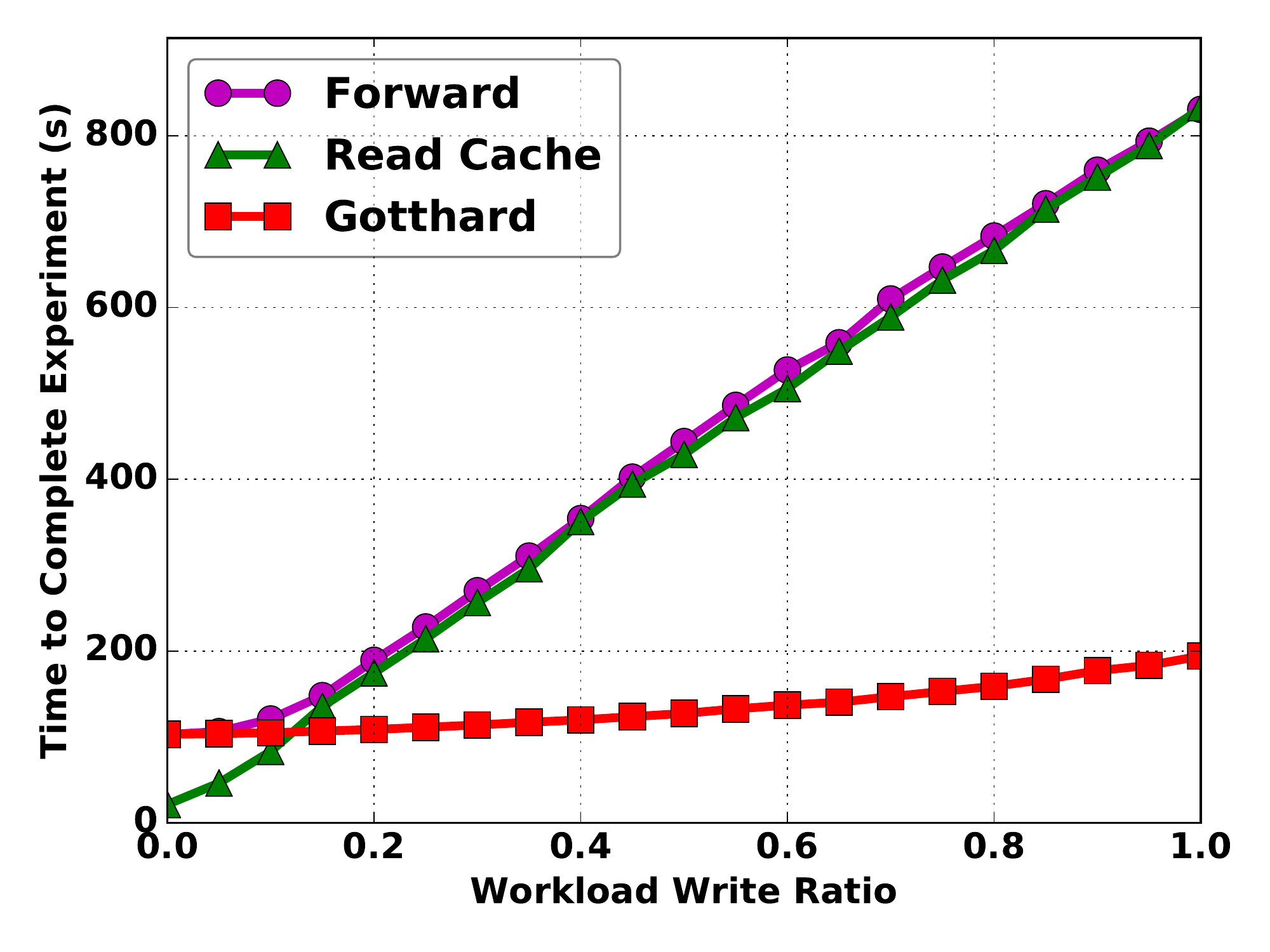}
        \caption{Time to complete 1000 transactions as the workload becomes more
          write-heavy. The read cache is ineffective as the percentage of write
          transactions exceeds 15\%.}
        \label{fig:motivation}
    \end{minipage}
\end{figure}

We consider a geographically distributed system composed of \emph{client} and \emph{server} processes. 
Processes communicate through message passing and do not have access to a shared memory. 
The system is asynchronous: there is no bound on messages delays and on relative process speeds. 
Processes are subject to crash failures and do not behave maliciously (e.g., no Byzantine failures). 

The store contains a set $D = \{ x_1, x_2, ... \}$ of data items.  Each data
item $x$ is a tuple $\langle k, v \rangle$, where $k$ is a key and $v$ a value.
We assume that the store exposes an interface with two operations:
\emph{read(k)} returns the value of a given $k$, and \emph{write(k,v)} sets
the value of key $k$ to value $v$.
We refer to those transactions that contain only read operations as
\textit{read transactions}.  Transactions that contain at least one write
operation are called \emph{write transactions}.

We assume that clients execute transactions locally and then submit the
transaction to the store to be committed.  When executing a transaction, the
client may read values from their own local cache. Write operations are buffered until commit time.

The isolation property that the system provides is \emph{one-copy
serializability}: every concurrent execution of committed transactions is
equivalent to a serial execution involving the same
transactions~\cite{bernstein87}.  

To ensure consistency, the store implements optimistic concurrency control
(OCC)~\cite{kung81}.  All \textit{read transactions} are served directly by the
store. To commit a \emph{write transaction}, the client submits its buffered
writes together with all values that it has read. The store only commits a
transaction if all values in the submitted transaction are still current.  As a
mechanism for implementing this check, the system uses a \emph{compare(k,v)}
operation, which returns true is the current value of $k$ is $v$, and false
otherwise. Note that the compare operation is not exposed to the users, but is
simply used by the system to implement optimistic concurrency control.
In the event of an abort, the server returns updated values, allowing the client to immediately re-execute the transaction.

\subsection{Motivation}

Figure~\ref{fig:cache} shows a typical deployment for a distributed
transactional storage system. Clients submit transactions to a store by sending
messages through the network.

The illustration shows a single switch in the network to
simplify the presentation, but in practice, the network topology can
be large and quite complex.  If the distance between the client and
the store is large, then the latency incurred by transmitting packets
over the network can significantly impact the system performance. To
reduce this overhead, systems typically deploy a
cache~\cite{freedman04,akamai}. The cache stores copies of some
subset of the data at the store. Client messages are routed to
the cache. If the cache has a copy of the requested data, then it can
respond to the clients request. Otherwise, it forwards messages
to the store.

As long as the cache is closer to the client than the store, then caching is an
effective technique for workloads that contain mostly read requests.  However,
caching does little to help improve performance for write-heavy workloads, since
write operations need to be forwarded to the store in order to maintain strong
consistency.

To demonstrate this behavior, we performed a simple experiment in which we
measured the total time to complete 1,000 transactions for increasingly
write-heavy workloads. In the experiment, a client submits two types of
transactions to the store: one with a single read operation, and one with both a
read and write operation. We varied the proportion of the two transaction types,
to create workload scenarios ranging from more read-intensive to more
write-intensive.

All the transactions passed through a switch that operated in one of three
different modes of execution. In the first, the switch acted traditionally, and
simply forwarded requests to the store. In the second mode, the switch was
modified using a data plane P4 program to behave like a cache. In
Figure~\ref{fig:cache}, the proxy cache is illustrated with a dashed-line to
indicate that in our experiment, the cache operations are actually executed by
the switch.  This is therefore an ``ideal'' cache, which has no network latency
between the switch and the proxy server hosting the cache. In the third
configuration, which will be explained in Section~\ref{sec:design}, the switch
executed Gotthard logic.

The experiment was run in emulation using the hardware described in
Section~\ref{sec:evaluation}. Figure~\ref{fig:motivation} shows the results. We
see that as the percentage of write requests increases, the cache becomes less
effective. In fact, when the percentage of write requests is above 15\%, we see
almost no benefit to performance for using the read cache. This is because the client can read stale
values from the cache which then cause write transactions to abort at the store.

As a preview of our results, Figure~\ref{fig:motivation} also includes
measurements for Gotthard which show how it improves performance for
write-intensive workloads.  The experiment demonstrates that Gotthard
significantly reduces execution time with respect to simple forwarding
and caching. When the workload is only at only 25\% writes, Gotthard
already reduces the completion time by half (i.e., a 2x improvement).

\section{System Design}
\label{sec:design}

\begin{figure}[t]
\centering
 \includegraphics[width=0.5\textwidth]{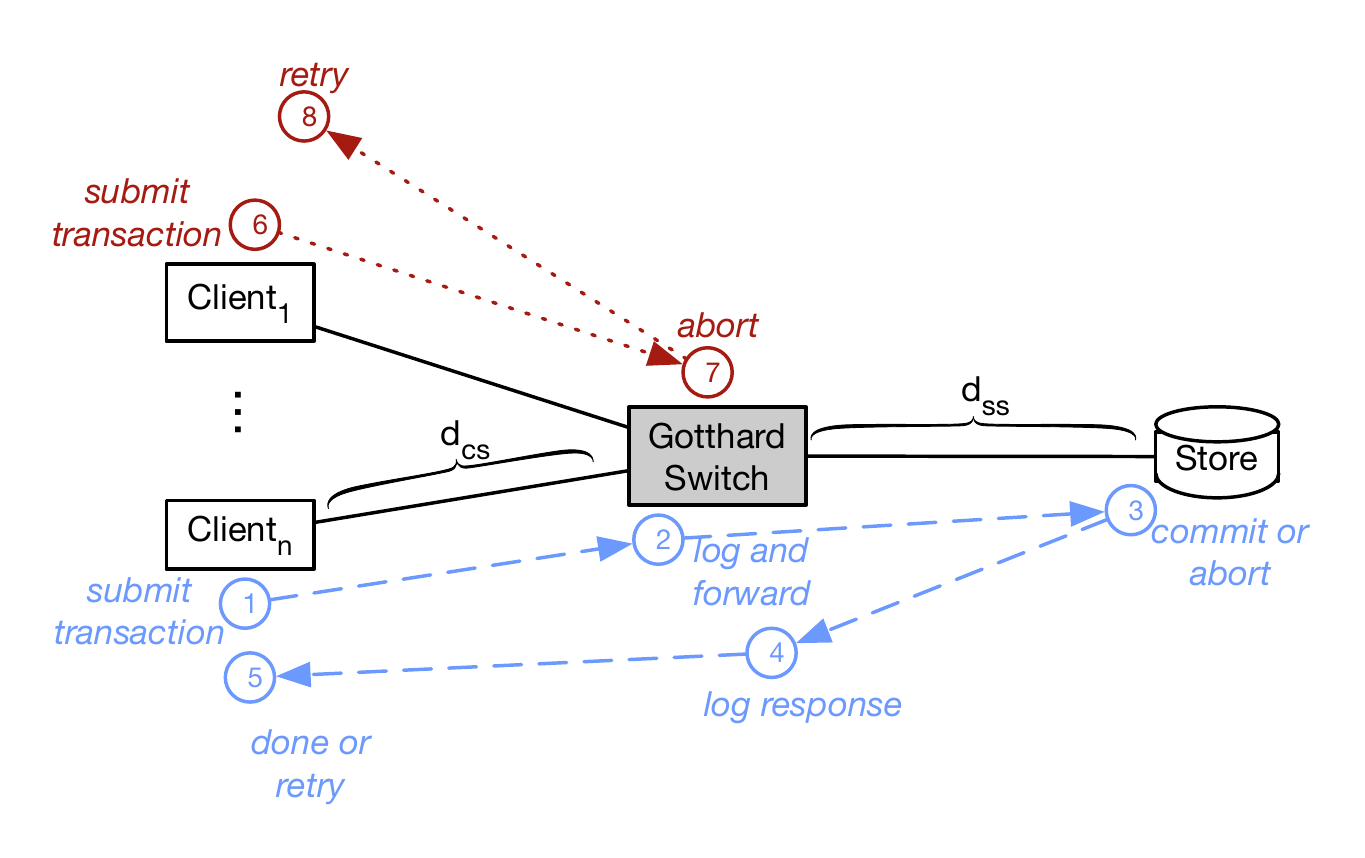}
\caption{Overview of Gotthard deployment.}
\label{fig:overview}
\end{figure}

It is widely recognized that the performance of optimistic concurrency control
protocols are heavily dependent on workload contention.  As contention
increases, so does the number of aborted transactions, causing the overall
performance of the system to drop~\cite{gruber97}.  Gotthard is designed to
address this problem.

In some respects, one can view Gotthard as providing complimentary behavior to read caches. For example,
SwitchKV~\cite{li16} forwards all transactions that contain only write
operations to the store, and possibly services read operations at the
cache. In contrast, Gotthard improves performance by
focusing on write-intensive workloads. 

However, although the approaches are complimentary, combining the two techniques
is not necessarily straightforward. Because all read operations in Gotthard are
serviced by the store, Gotthard enforces single-copy serializability (a
proof-sketch of the correctness is in Section~\ref{sec:correctness}). Servicing read operations at a proxy server, or even a switch, would
require additional mechanisms to ensure that the cache entries are valid.

\subsection{Overview}

Figure~\ref{fig:overview} shows a basic overview of Gotthard. From a high-level,
clients submit transaction requests to the store. The requests pass through a
Gotthard switch, which either forwards the request on to the store, or aborts
the transaction, and responds to the client directly. In the figure, the two
cases are distinguished by color and line type.

The blue, dashed-line shows the forwarding case. When the client submits the
request (1), the switch examines the transaction and logs the operations in its
local cache (2). It then forwards the transaction message to the store, which
can commit or abort the transaction (3). The store responds to the client with
the outcome of the execution. The switch logs the result of the execution (4),
and forwards the response to the client. If the client learns that the
transaction was aborted, it can re-try. Otherwise, the transaction is complete (5).

The red, dotted-line shows the abort case. As before, the client submits a
request (6), and the switch examines the transaction message. When logging the
operations, if the switch sees that a transaction is likely to abort based on
some previously seen transaction, the switch will preemptively abort the request
(7), and send a response to the client. The client can then re-submit the
transaction (8).

For a transaction that would have aborted at the store, the intuitive advantage
of the Gotthard approach is clear: the message avoids traveling the distance
from the switch to the store, $d_{ss}$, twice. However, within this basic
framework, there are a number of subtle design decisions that impact the
performance. We discuss these in more detail below.

Note that in the following discussion, we refer to the state that Gotthard uses
to log transactions as a cache. However, to be clear, Gotthard only uses this
state to make decisions about aborting transactions. It does not service read
requests from its cache. As we discuss in the next section, the values that
Gotthard keeps in its cache may be stale or incorrect.

In most of the following text, we refer to a basic deployment with a single
switch to make exposition simpler. In Section~\ref{sec:deployment}, we discuss
more complex topologies.

\subsection{Data Store}

\begin{algorithm}[t]
\SetKwBlock{Upon}{Upon}{}
\DontPrintSemicolon % Some LaTeX compilers require you to use \dontprintsemicolon instead

\Upon(receiving TXN{(}\textit{compares, reads, writes}{)}:\label{algo:store:txn}){
     \textit{corrections} $\gets$ $\emptyset$\;
     \ForEach{(key, value) $\in$ compares}{
         \uIf{value $\neq$ store[key]}{
             \textit{corrections} $\gets$ \textit{corrections} $\cup$ (\textit{key}, \textit{store}[\textit{key}])
             }
         }
    \lIf{corrections $\neq$ $\emptyset$}{
        respond Abort(\textit{corrections})\label{algo:store:abort}
    }
    \ForEach{(key, value) $\in$ writes}{
        \textit{store}[\textit{key}] $\gets$ \textit{value}\label{algo:store:write}
    }
    \textit{response} $\gets$ \textit{writes}\;
    \ForEach{key $\in$ reads}{
        \textit{response} $\gets$ \textit{response} $\cup$ (\textit{key}, \textit{store}[\textit{key}])\label{algo:store:read}
    }
    respond OK(\textit{response})\;
}
\caption{Store Transaction Processing}
\label{algo:store}
\end{algorithm}

Algorithm~\ref{algo:store} shows the logic executed by the store.  A transaction
request message (\texttt{TXN}) contains three possibly empty lists of
operations: \emph{compares}, \emph{reads}, and \emph{writes}.  The store first
iterates over the compare operations to check for stale values (i.e., the value
of the compare operation does not match the current value in the store). If any
comparison operation fails, the store aborts the transaction
(line~\ref{algo:store:abort}). As part of the abort response, the store includes
a list of correct values (\texttt{corrections}), with the updated values for
comparisons that caused the transaction to fail. Otherwise, the store applies any write
operations the transaction may include to update the values of the data items
(line~\ref{algo:store:write}). Then the store responds to the client with all
the values that were updated, along with the values that the transaction reads
(line~\ref{algo:store:read}).

\subsection{Optimistic Abort}

\begin{algorithm}[t]
\SetKwBlock{Upon}{Upon}{}
\DontPrintSemicolon % Some LaTeX compilers require you to use \dontprintsemicolon instead

\Upon(receiving TXN{(}\textit{compares, reads, writes}{)} \textbf{from} client:\label{algo:optiabort:client}){
    \textit{corrections} $\gets$ $\emptyset$\;
    \ForEach{(key, value) $\in$ compares}{
        \uIf{key $\not\in$ cache}{\label{algo:optiabort:miss}
            forward to store\label{algo:optiabort:forward1}
        }
        \uElseIf{value $\neq$ cache[key]}{
            \textit{corrections} $\gets$ \textit{corrections} $\cup$ (\textit{key}, \textit{cache}[\textit{key}])\label{algo:optiabort:corr}
        }
    }
    \uIf{corrections $\neq$ $\emptyset$}{
        respond Abort(\textit{corrections})\label{algo:optiabort:abort}
    }
    \uElseIf{reads $\cup$ writes $\neq$ $\emptyset$}{
        \ForEach{(key, value) $\in$ writes}{
            \textit{cache}[\textit{key}] $\gets$ \textit{value}\label{algo:optiabort:update}
        }
        forward to store\label{algo:optiabort:forward2}
    }
    \lElse{respond OK()\label{algo:optiabort:ok}}
}
\Upon(receiving Abort{(}\textit{corrections}{)} \textbf{from} store:\label{algo:optiabort:store}){
    \ForEach{(key, value) $\in$ corrections}{
        \textit{cache}[\textit{key}] $\gets$ \textit{value}\label{algo:optiabort:store2}
    }
    forward to client
}
\caption{Gotthard Switch Logic}
\label{algo:optiabort}
\end{algorithm}

A Gotthard switch aggregates messages from several clients. The switch logs
requests and responses in order to determine if a subsequent transaction is
likely to abort. Gotthard adopts an aggressive strategy for aborting
transactions. It proactively updates its cache with the latest value after
the switch has seen a transaction request (step 2 in Figure~\ref{fig:overview}).

We refer to this as an \emph{optimistic abort} strategy. It is optimistic
because the switch assumes that any transaction request that is has seen is
likely to be committed. As a result, it can make decisions about aborting
subsequent transactions sooner. However, this approach may abort transactions
that would not have been aborted by the store.  

The logic for early optimistic abort is presented in
Algorithm~\ref{algo:optiabort}. The switch has logic for processing both transaction requests
(line~\ref{algo:optiabort:client}) and their responses
(line~\ref{algo:optiabort:store}).  When the switch receives a transaction from the client, it iterates over the
list of compare operations. If any compare operation references a key that is
not in the cache (line~\ref{algo:optiabort:miss}), then the switch cannot reason
about the validity of the transaction, so it forwards the request to the store
(line~\ref{algo:optiabort:forward1}). If the compare operation references a key
that is in the cache, then the switch compares the value in the packet, with
that in the store. If the values differ, the current value in the cache is added to a
per-transaction set of corrections (line~\ref{algo:optiabort:corr}). After
processing the compares, if there is
at least one correction (i.e. there was a comparison that failed), the switch
immediately sends an abort response to the client with the list of corrections
(line~\ref{algo:optiabort:abort}).

If no comparisons failed, then they switch checks if the request contains
\emph{read} or \emph{write} operations. If there are write operations, then it
updates its cache with the new values (line~\ref{algo:optiabort:update}). Then, it forwards the transaction
to the store for processing. (line~\ref{algo:optiabort:forward2}). Otherwise, if the transaction only
included compares, the switch returns a successful response  immediately to the client
(line~\ref{algo:optiabort:ok}).

An abort message from the store (\texttt{Abort}) contains a non-empty list of the
corrections. The corrections contain the updated values that caused the
transaction to fail.  When the switch receives an abort message
(line~\ref{algo:optiabort:store}), it updates its cache with the correct values
(line~\ref{algo:optiabort:store2}). If the switch did not update its cache on
aborts, it would incorrectly abort subsequent transactions.

We note that although a Gotthard switch needs to maintain state in its local
cache, the size of the cache does not need to be too large to be
effective. It is sufficient to reserve enough space to keep for ``hot'' data
items. The amount of space available for the cache will depend on the target
platform for deployment. If the size is restricted, Gotthard could use a
least-recently used cache eviction policy to make space available for new items.

\subsection{Version Numbers}

 In a typical transactional storage system, data items 
 would include a version number.  In other words, each data item $x$
would be a tuple $\langle k, v, ts \rangle$, where $k$ is a key, $v$ its value,
and $ts$ its version. The store would use the version numbers to determine if a
transaction should be aborted due to a stale value.
Notably, with this design, the store is responsible for assigning version
number to data items. 

However, with the optimistic abort strategy, the switch
must update its local cache of data items before the store can assign a
version number. Therefore,  Gotthard
cannot use version numbers to check for stale values. 
Instead, both the switch and the store compare with the current value to see if
the data item for a specific key has changed.

\subsection{Gotthard Messages}
\label{sec:gotthardmsgs}

All transactions and transaction responses are encoded in a custom Gotthard
packet header. Gotthard headers are encapsulated inside UDP transport-layer protocol headers.
Prior work has shown that UDP is suitable for high-performance key-store
systems~\cite{nishtala2013scaling}.  If the set of operations in a transaction
exceed the maximum transmission unit (MTU), Gotthard transactions may be 
 split across multiple packets.

The header is shown in P4 syntax in
Figure~\ref{fig:header}. The transaction header, \texttt{gotthard\_hdr\_t}, contains eight fields:

\begin{itemize}
\item \texttt{msg\_type} indicates is the message is a request to or a response
  from the store.
\item \texttt{from\_switch} is a flag that indicates if an abort was performed
  by the switch or the store. This is currently used in our experiments for
  statistical purposes.
\item \texttt{txn\_id} is a unique identifier for each transaction.
\item \texttt{frag\_count} is the total number of fragments, and is used if a
  set of operations exceeds the packet maximum transmission unit (MTU).
\item \texttt{frag\_seq} identifies which fragment out of the total this packet contains.
\item \texttt{status} indicates a commit or abort. 
\item \texttt{op\_cnt} is the number of operations in the transaction.
\end{itemize}

\noindent
The transaction header is followed by an array of operations. Each operations is
defined in P4 as a separate header, \texttt{gotthard\_op\_t}. The
\texttt{op\_type} indicates the type of operation (compare, read, or write) and
the \texttt{key} and \texttt{value} are the operation operands. For read
operations, the value operand is unused. For compare operations, the value
operand contains the value to compare.

An alternative implementation could have used different headers for transaction
requests and responses. We chose this implementation because it simplified the P4
logic for aborting a message. The switch simply needs to overwrite the compare
fields with the correct values, and change the \texttt{msg\_type}. In contrast,
removing and then appending a new header would have been an expensive operation
in the software implementation of the P4 switch.

\subsection{Expected Deployment}
\label{sec:deployment}

As our evaluation in Section~\ref{sec:evaluation} will show, Gotthard is most
effective when the Gotthard switch is (i) deployed close to clients, and (ii)
when the workload exhibits a high-degree of write contention. 

\begin{figure}[t]
    \centering
    \begin{minipage}{.45\textwidth}
        \centering
         \includegraphics[width=\textwidth]{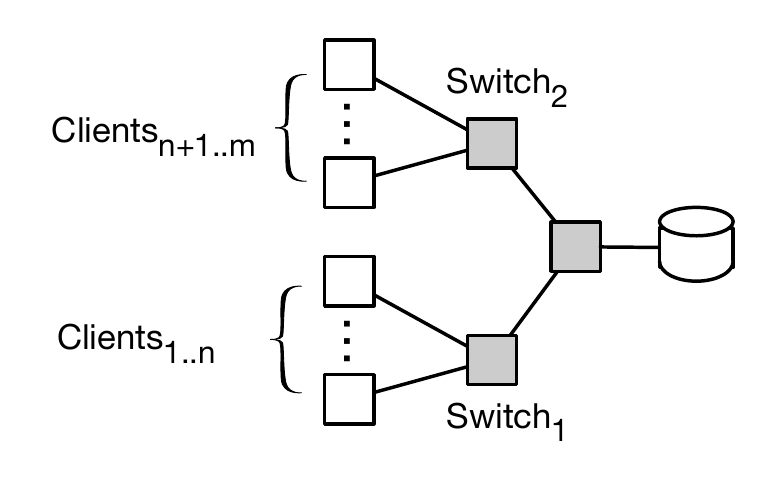}
        \caption{Gotthard is most effective when clients that connect to the same switch 
         exhibit a high-degree of write contention.}
        \label{fig:locality}
    \end{minipage}%
    \hfill
    \begin{minipage}{.45\textwidth}
        \centering
\begin{lstlisting}[xleftmargin=5.0ex]
header_type gotthard_hdr_t {
  fields {
    bit<1> msg_type;    // REQ or RES
    bit<1> from_switch;
    bit<32> txn_id;
    bit<8> frag_seq;
    bit<8> frag_cnt;
    bit<8> status;
    bit<8> op_cnt;
    // followed by op_cnt gotthard_op_t headers
  }
}

header_type gotthard_op_t {
  fields {
    bit<8> op_type;
    bit<32> key;
    bit<1024> value;
  }
}
\end{lstlisting}
        \caption{Gotthard packet header in P4 syntax.}
        \label{fig:header}
    \end{minipage}
\end{figure}

An ideal deployment for Gotthard is one in which client locality correlates with
write contention. For example, in Figure~\ref{fig:locality}, we assume that
clients $1 \dots n$ frequently access the same data items, and clients $(n+1)
\dots m$ access a different set of data items. Two switches, (Switch 1
and 2), are deployed close to their respective clients, and examine and
potentially abort writes from the different sets of data items.

Such partitioning of data and clients occurs naturally in a number of
applications. For example, if the clients are sensors, they are likely to read
data from overlapping sets of items. In the TPC-C benchmarks that we used in our
evaluation, data is naturally partitioned by clients accessing different
warehouses that are geographically distributed.

We should note that if clients connected to different switches update data at
the store, the correctness of Gotthard is not violated. Clients and switches
will learn of new values after an abort message from the store. For example, if client$_1$
writes a value $v_1$ for key $k_1$, then switch$_1$ will record $v_1$ in its
cache. However, if client$_{n+1}$ had previously written a value
$v_1'$ for key $k_1$,  the request from client$_1$ will pass through switch$_1$,
but will be aborted by the store. The client and switch$_1$ will learn of of the
new value $v_1'$  in the abort response from the store. They will both then
update their local caches, and the client can re-submit the transaction with the
latest value. 

\subsection{Correctness}
\label{sec:correctness}

The storage (Algorithm~\ref{algo:store}) ensures serializable executions by
means of optimistic concurrency control.  The serialization order is defined by
the order transactions arrive at the store.  A transaction only commits if all
the reads it performed during execution are still up to date at the time the
transaction is received by the store.

The correctness of the switch logic (Algorithm~\ref{algo:optiabort}) follows
from the fact that (a)~the switch does not commit any transactions, although it
may abort transactions, and (b)~the switch forwards non-aborted transactions to
the store without changing their operations.

\subsection{Fault Tolerance} 

It is often desirable for storage systems to provide fault tolerance.
Fault-tolerance is an orthogonal problem to those that Gotthard
addresses. However, the system could address failures by replicating the store
using any of the standard consensus
protocols~\cite{lamport98,OL88,ongaro14,CT96}, and adding rules to the switch to
route traffic to a replica in the event of failure (e.g., similar to OpenFlow's
fast fail-over).

\section{Implementation}
\label{sec:implementation}

We have implemented a prototype of Gotthard. The switch logic is written in the
P4 programming language~\cite{bosshart14}, version 1.10~\cite{p4spec} (i.e.,
P4-14), although porting to P4-16 should be straightforward.  We 
used P4.org's reference compiler~\cite{p4org} targeting the Behavioral Model
switch~\cite{bm}. The client and store code is written as standalone Python
programs, using Python 2.7.

To be sent over Ethernet links, the size of the Gotthard message cannot exceed
the Ethernet MTU of 1,500 bytes, otherwise the message will be fragmented at the
network level. If a transaction has more than 10 operations, the message would
exceed the MTU, so we fragment the Gotthard message at the application level
using the header fields described in Section~\ref{sec:gotthardmsgs}. When the
Gotthard switch receives a fragmented message, it does execute any of
Algorithm~\ref{algo:optiabort}, but just forwards the message. This is because
it is not possible to reason about the validity of a transaction without seeing
it in its entirety.

All code, including the switch, client, store, simulation and experiment harness
is publicly available with an open-source
license.\footnote{\url{https://github.com/usi-systems/gotthard}}

\section{Evaluation}
\label{sec:evaluation}

In this section, we describe two sets of experiments that evaluate Gotthard. The
first set of experiments are microbenchmarks that explore how Gotthard impacts
the performance for executing transactions on a key-value store under various
operational conditions, including relative distance of the switch to clients and the store,
workload contention, and data locality on a multiple switch setup. In the second set of
experiments, we explore the effectiveness of Gotthard for improving the performance
for real-world inspired workloads, by using the TPC-C~\cite{tpcc} benchmark with
the parameters adjusted to increase contention.
We ran our experiments using both a network emulator for a controllable
environment, and on Amazon's EC2~\cite{ec2} for a more realistic setting using virtual
machines in the cloud. Overall, the results demonstrate that Gotthard
significantly reduces latency and improves throughput for workloads with high contention.

\subsection{Experimental Setup}
\label{sec:evaluation:setup}

\begin{table}[t]
    \centering
    \begin{minipage}{.45\textwidth}
        \renewcommand{\arraystretch}{1.2}
        \centering
        \ra{1.3}
         \begin{tabular}{ @{}cccccc@{} }
         \hline
         \toprule
                    & \textbf{VA} & \textbf{EU} & \textbf{OR}  & \textbf{JP} \\
          \midrule
        \textbf{CA} & 70-79 & 143-148 & 20-22    & 102-113 \\
        \textbf{VA} & -     & 70-76   & 66-85    & 140-153 \\
        \textbf{EU} & -     & -       & 126-131  & 209-218 \\
        \textbf{OR} & -     & -       & -        & 89-100 \\
         \bottomrule
         \end{tabular}
         \caption{RTT between AWS regions (milliseconds)}
        \label{tab:awsrtt}
    \end{minipage}%
    \hfill
    \begin{minipage}{.45\textwidth}
\renewcommand{\arraystretch}{1.2}
\centering
\ra{1.3}
 \begin{tabular}{ @{}ll@{} }
 \hline
 \toprule
\textbf{Parameter} & \textbf{Default}  \\
\midrule
RTT & 100 ms \\ 
delta & 0.2 \\
\#clients & 8 \\
\%writes & 0.2 \\
locality & 1.0 \\
 \bottomrule
 \end{tabular}
 \caption{Default parameter values in microbenchmarks.}
\label{tab:defaults}
    \end{minipage}
\end{table}

For all emulator-based experiments, we used
Mininet~\cite{mininet} to create a virtual network. All experiments ran 
on a server machine with 12 cores (dual-socket Intel Xeon E5-2603 CPUs @ 1.6GHz)
and 16GB of 1600MHz DDR4 memory. The operating system was Ubuntu 14.04.

We ran all cloud based experiments using Amazon's EC2~\cite{ec2},
deploying on EC2 instances of type \texttt{t2.medium}.  The clients,
switch, and store were placed in individual instances, located in
geographically distributed regions.  Specifically, the clients, switch and store
were placed in regions \texttt{us-west-2}, \texttt{us-west-1} and
\texttt{us-east-1}, respectively.  We observed RTT latencies of 20-25ms between the clients and
the switch, and 60-82ms between the switch and the store.

In both the emulator and EC2 experiments, we used the P4 Behavioral Model
switch~\cite{bm} to run the Gotthard P4 program. The same Python code was
used in both deployments for the client and store.

In all experiments, we compare the performance of three different approaches:
\begin{itemize}
\item \emph{Forwarding.} In this deployment, the switch simply forwards
  transaction messages to the intended destination, without executing any
  logic. This approach represents a baseline deployment.
\item \emph{Read Cache.} This deployment represents an idealized on-path
    look-through read cache deployment in which the switch itself caches values,
        and services read requests if the data is in the cache.
\item \emph{Gotthard.} The switch that executes the transaction logic described in
  this paper.
\end{itemize}

Except for the multi-switch microbenchmark, we used a topology in which all
messages to and from the store are routed through a single switch, as shown in
Figure~\ref{fig:cache}.

\subsection{Calibration}

Gotthard is designed to reduce the overhead due to network latency for
transaction processing. Network latency is a particular concern for wide area
and geo-distributed deployments. In order to calibrate the expected network
latencies in our simulations, we first measure round-trip time (RTT) latencies
one might expect for a global deployment on EC2.

EC2 is hosted in multiple global regions. Each region corresponds to a
geographic location and is divided into
multiple availability zones.  We started a \texttt{m3.large} VM
instance in each availability zone, and ran the ping client to measure RTT to each other running instance.
Measurements were collected for 2 minutes. Table~\ref{tab:awsrtt} reports the minimum
and maximum average for each region. The intra-region latencies were between 0.8
and 1.2 milliseconds, except for the JP region, where it was 2.5
milliseconds.

These measurements identify the range of network latencies one can expect in
geo-distributed deployments. In all microbenchmark experiments in emulation, we
used a fixed RTT of 100 milliseconds (i.e., 50 milliseconds one way).  This
number is well within the range of latencies one would expect on EC2 and is
close the latencies observed in our actual EC2 experiments.

\subsection{Microbenchmarks}

To understand the effect of specific operational conditions on the performance of Gotthard, we designed a set of microbenchmarks, each focusing on one the following parameters:
\begin{itemize}
\item \emph{Delta ratio.} The delta ratio captures the relative distance of the
  switch to clients and the store. Using the variables in
  Figure~\ref{fig:overview}, the delta ratio is defined as $d_{cs} / d_{cs} +
  d_{ss}$. So, a smaller ratio means that the switch is closer to the
        client. Note that  $d_{cs} + d_{ss} = \frac{RTT}{2} = 50 \mbox{ms}$.
\item \emph{Percentage of writes.} Each client in the benchmarks issues a mix of \textit{read} and \textit{write} transactions. Percentage of writes dictates how many of the total number of transactions are writes.
\item \emph{Number of clients.} The number of concurrent clients submitting transactions.
\item \emph{Contention.} To characterize contention, we model the popularity of keys in the store using a Zipf distribution. Contention is increased by increasing the Zipf exponent; when the exponent is 0, all keys are equally likely to be accessed by clients.
\item \emph{Locality on a multi-switch deployment.} In a multi-switch setup such
  as the one described in Figure~\ref{fig:locality}, it is possible that the
  optimistic execution of Gotthard results in false positives; transactions
  built upon optimistic state are aborted at the store. How likely that is to
  happen will depend on the workload. \textit{Locality} describes the
  probability of clients accessing keys ``owned'' by another switch.
\end{itemize}

In the microbenchmarks, each transaction accesses a single key, representing a
counter.  A \textit{read} transaction reads the current counter value and a
\textit{write} transaction atomically increments a counter.

\begin{figure}[ht!]
\centering
   \begin{subfigure}[t]{.45\textwidth}
     \centering
      \includegraphics[width=\textwidth]{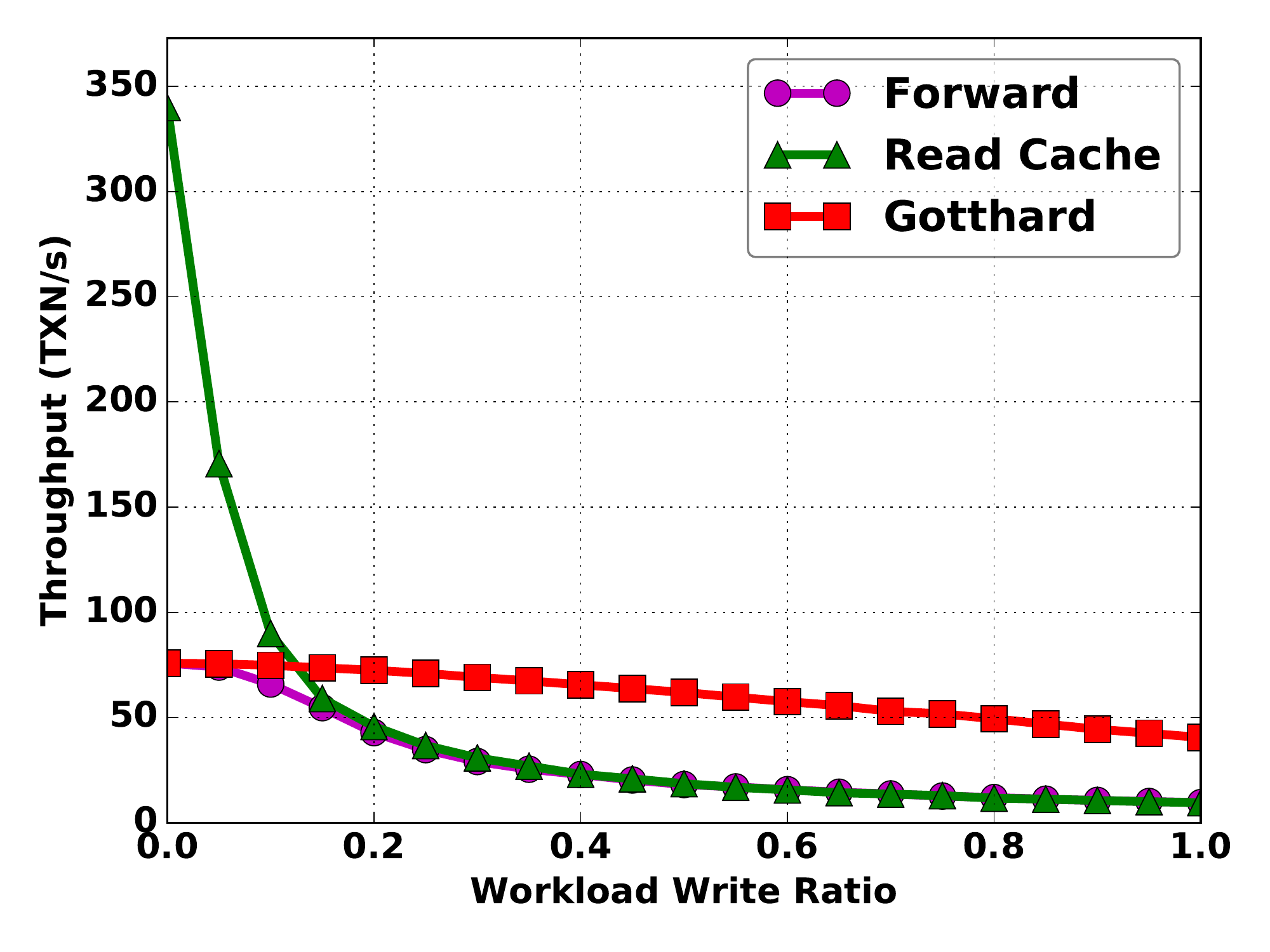}
      \caption{\% writes vs. throughput.}
      \label{fig:write-throughput}
    \end{subfigure}
\begin{subfigure}[t]{.45\textwidth}
     \centering
      \includegraphics[width=\textwidth]{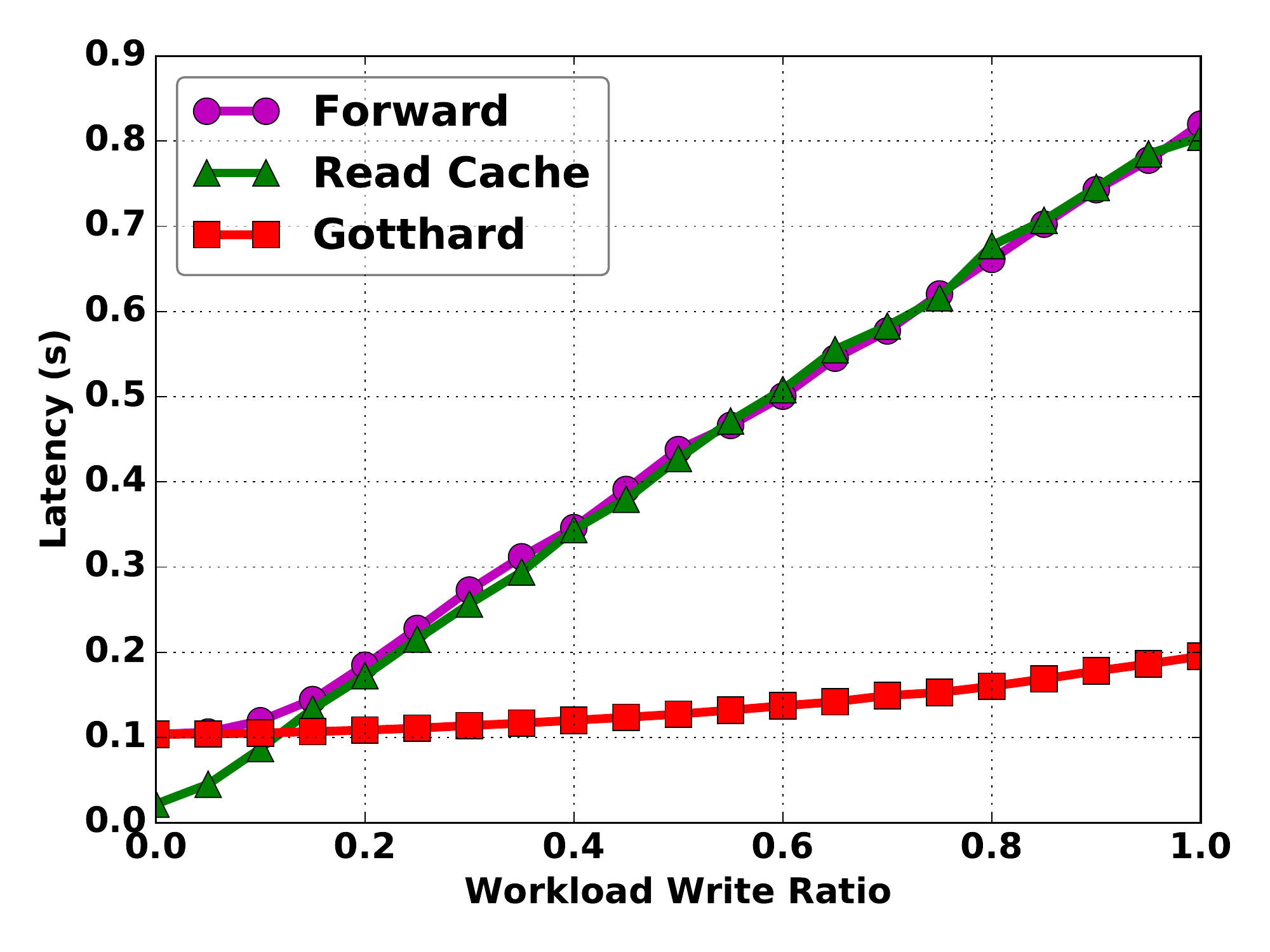}
      \caption{\% writes vs. latency.}
      \label{fig:write-latency}
    \end{subfigure}
\caption{Performance as \%writes changes.}
\end{figure}

For each experiment, unless otherwise specified, we used the default parameter
settings in Table~\ref{tab:defaults} and the topology in Figure~\ref{fig:overview}.
We deliberately chose a lower value of 0.2 \%writes to better understand the effect of other parameters on performance, but as Figure~\ref{fig:write-throughput} shows, Gotthard's relative performance improves with higher \%writes.
For each data point, we ran the experiment for three
minutes and report throughput in transactions committed per second, and the
average latency of committed transactions.

\begin{figure}[ht!]
\centering
  \begin{subfigure}[t]{.45\textwidth}
     \centering
     \includegraphics[width=\textwidth]{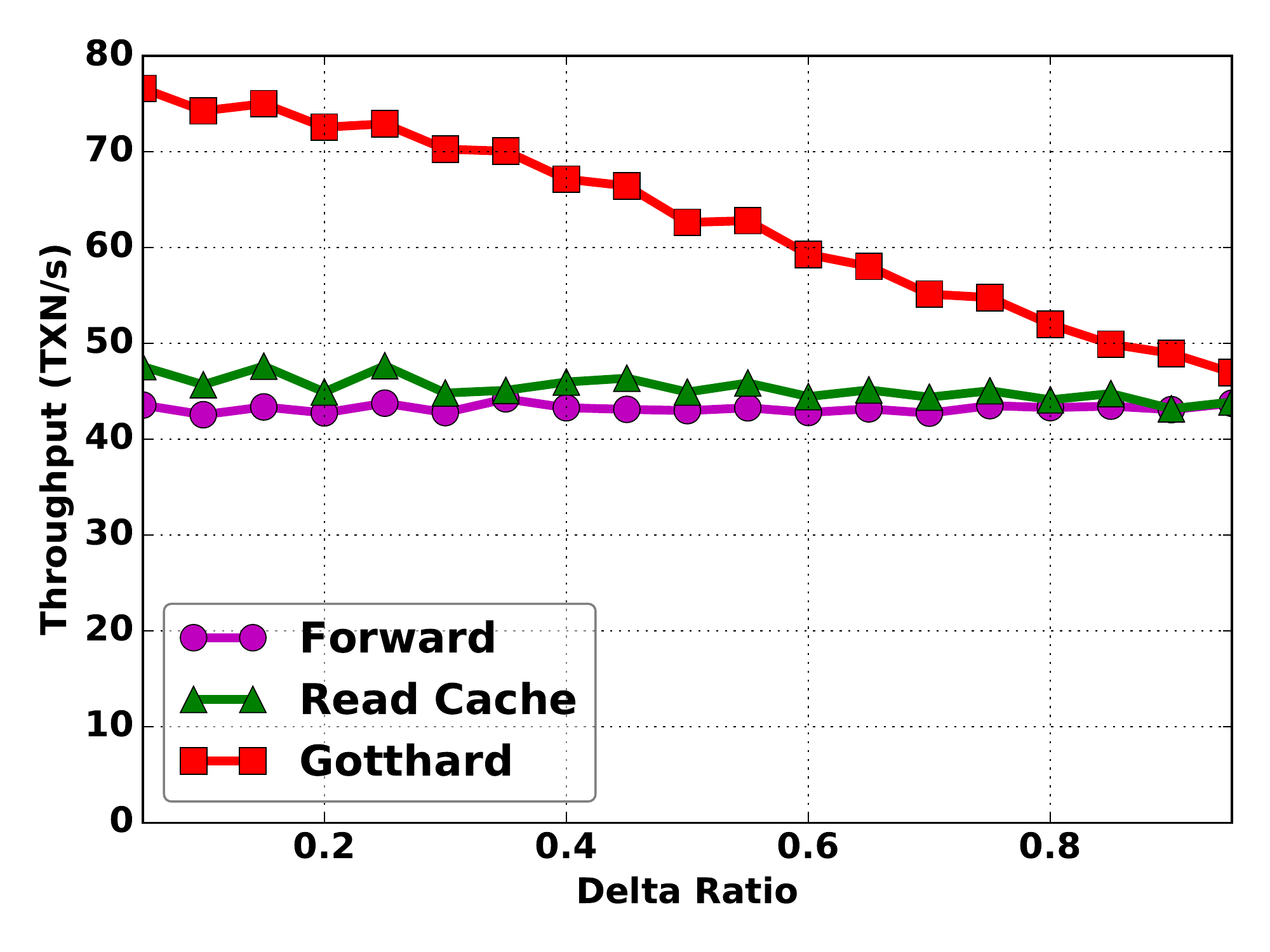}
     \caption{Delta ratio vs. throughput}
     \label{fig:delta-throughput}
   \end{subfigure}
\begin{subfigure}[t]{.45\textwidth}
     \centering
     \includegraphics[width=\textwidth]{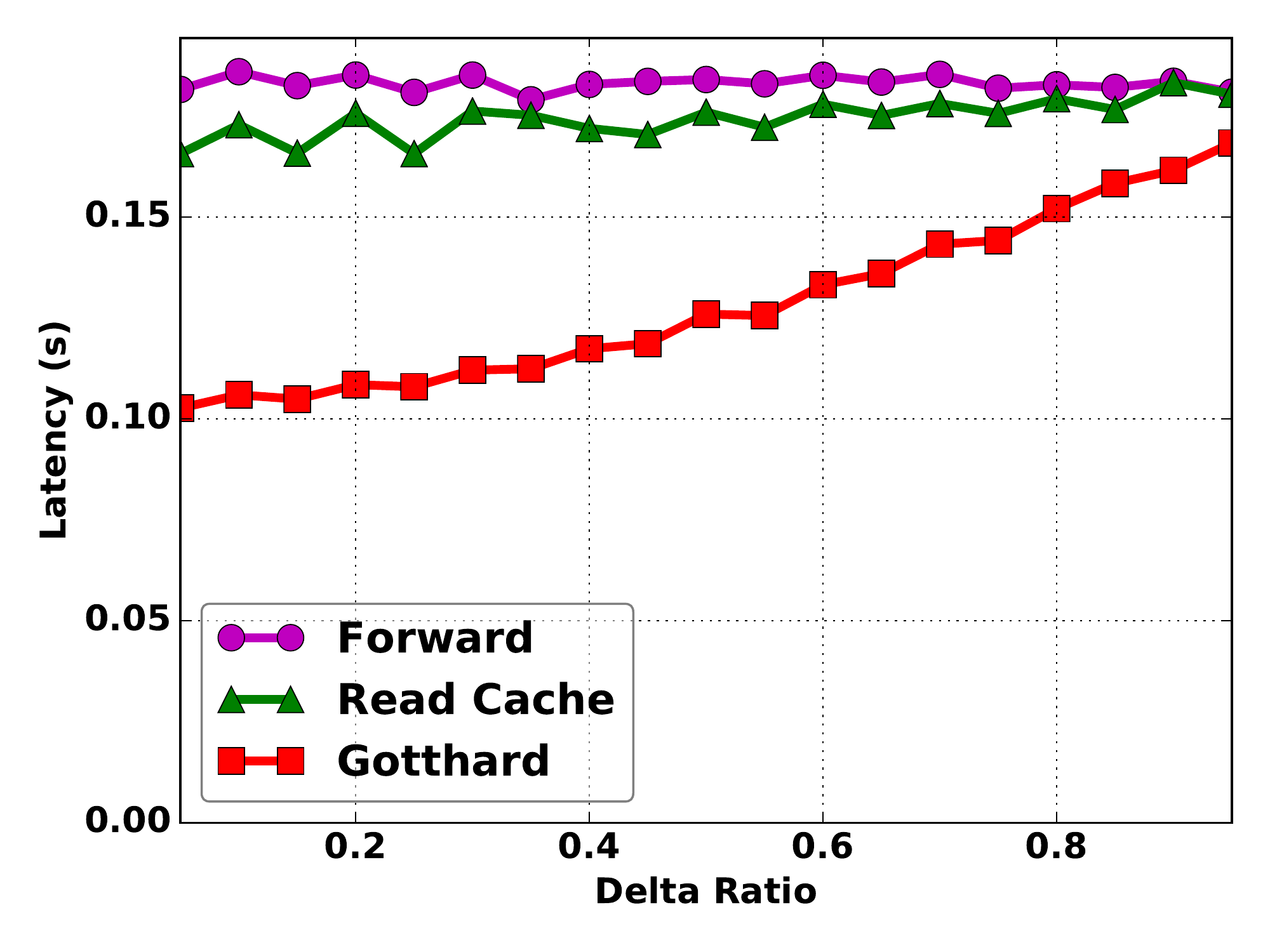}
     \caption{Delta ratio vs. latency}
     \label{fig:delta-latency}
   \end{subfigure}
\caption{Performance as delta ratio increases.}
\end{figure}

\paragraph*{Delta ratio.}
In this experiment, we quantify the effect of the relative distance of the
switch to clients and the store: the \textit{delta ratio}.  As the ratio
increases, the switch gets further from the client and closer to the store.
Figure~\ref{fig:delta-throughput} reports overall throughput as the delta ratio
changes from 0 to 1.  As expected, as the switch get closer to the client (i.e.,
delta approaches 0), the overall throughput for Gotthard improves because transactions
can be aborted and restarted earlier.  Even though the performance of the Read
Cache also improves slightly with a smaller delta ratio, its overall throughput
is limited by write transaction throughput, since
conflicting writes must reach the store to abort.
Figure~\ref{fig:delta-latency} reports average latency.  Gotthard's latency in
this scenario is always lower than the other approaches, and is dependent on the
latency between client and switch.  Under high contention, write latency
dominates the overall latency for Read Cache, due to costly aborts.

\paragraph*{Percentage of writes.}
Figures \ref{fig:write-throughput} and \ref{fig:write-latency} show
throughput and latency, respectively, as \%writes changes. These figures clearly demonstrate the
effect of write operations, which limit the overall performance of the system.  With
\%writes at 0, Read Cache can always serve requests from the switch, but as the
\%writes grows, throughput becomes limited due to the high cost of aborting
writes.  Gotthard's throughput, on the other hand, degrades slowly, since
requests are aborted at the switch and clients can optimistically retry
transactions sooner.  At 0.2 \%writes, Gotthard's throughput is already
1.5x that of Read Cache, reaching 3.3x that of Read Cache at around 0.5
\%writes.

\begin{figure}[ht!]
\centering
   \begin{subfigure}[t]{.45\textwidth}
     \centering
 \includegraphics[width=\textwidth]{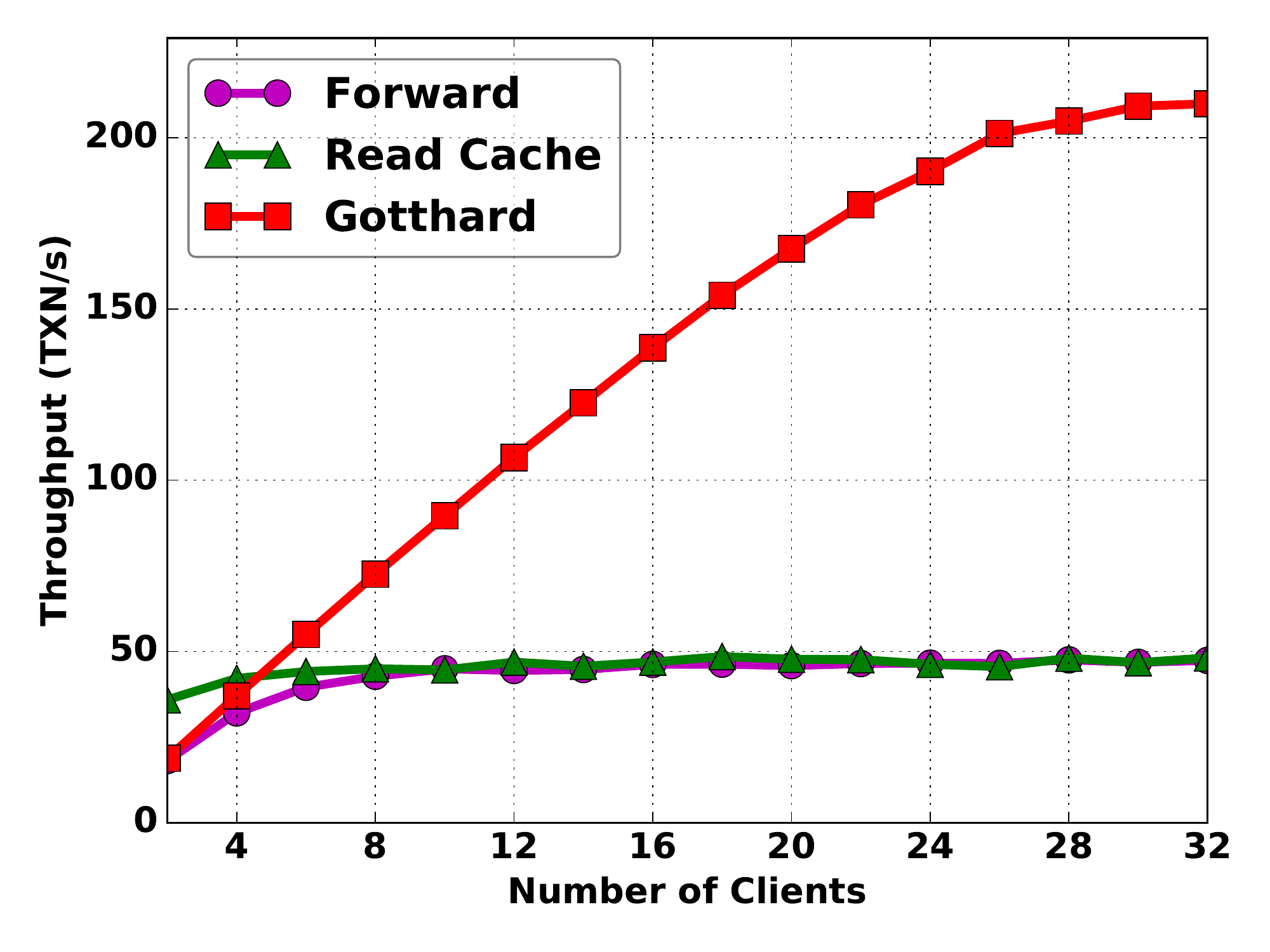}
 \caption{\#clients vs. throughput.}
 \label{fig:clients-throughput}
 \end{subfigure}
\begin{subfigure}[t]{.45\textwidth}
     \centering
 \includegraphics[width=\textwidth]{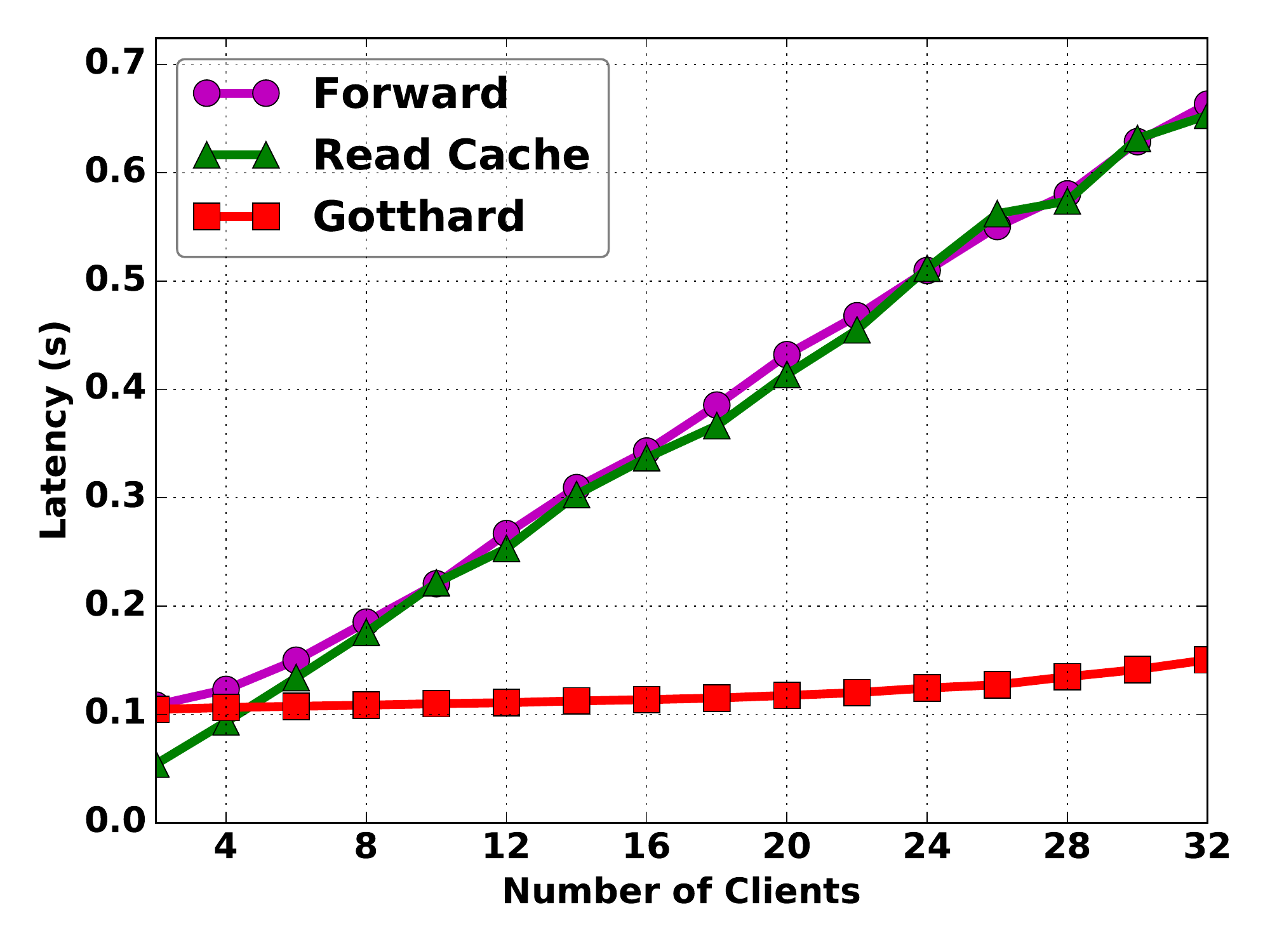}
 \caption{\#clients vs. latency.}
 \label{fig:clients-latency}
 \end{subfigure}\\
\caption{Performance as load (\#clients) increases.}
\end{figure}

\begin{figure}[ht!]
\vspace{-2mm}
\centering
 \includegraphics[width=0.45\textwidth]{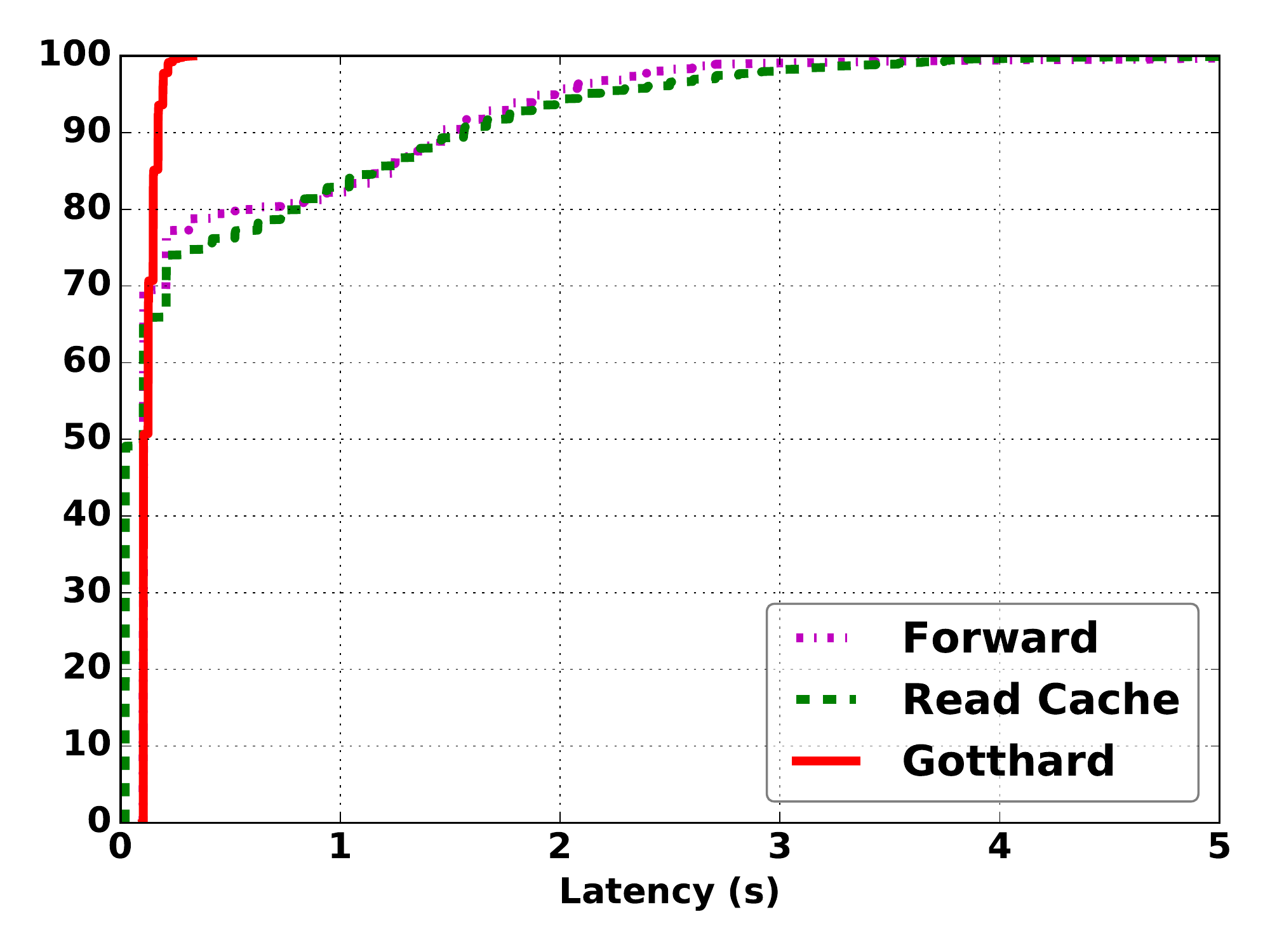}
\caption{Latency CDF for fixed delta, \%writes, clients.}
\vspace{-1mm}
\label{fig:cdf}
\vspace{-1mm}
\end{figure}

\paragraph*{Number of clients.}
Figure~\ref{fig:clients-throughput} reports throughput, and demonstrates how
Gotthard allows a higher rate of transactions under contention, past the
saturation point of the other two approaches, reaching more than 4x times the throughput of Read Cache.
Using Gotthard, transactions can abort early and optimistically be retried with updated values,
\textit{before the previous conflicting transaction commits}.  If the optimistic retries are
not aborted at the store, more contending transactions can be executed
concurrently.  Figure~\ref{fig:clients-latency} shows that Gotthard's latency
remains stable under increasing load, until it starts reaching saturation at
around 24 clients. Figure~\ref{fig:cdf} shows the CDF for latencies when load is fixed at 8 clients.
The ``fat'' tail observed for Read Cache and Forward are due to write transactions that repeatedly abort due to contention.

\begin{figure}[ht!]
    \centering
  \begin{subfigure}[t]{.45\textwidth}
    \centering
    \includegraphics[width=\textwidth]{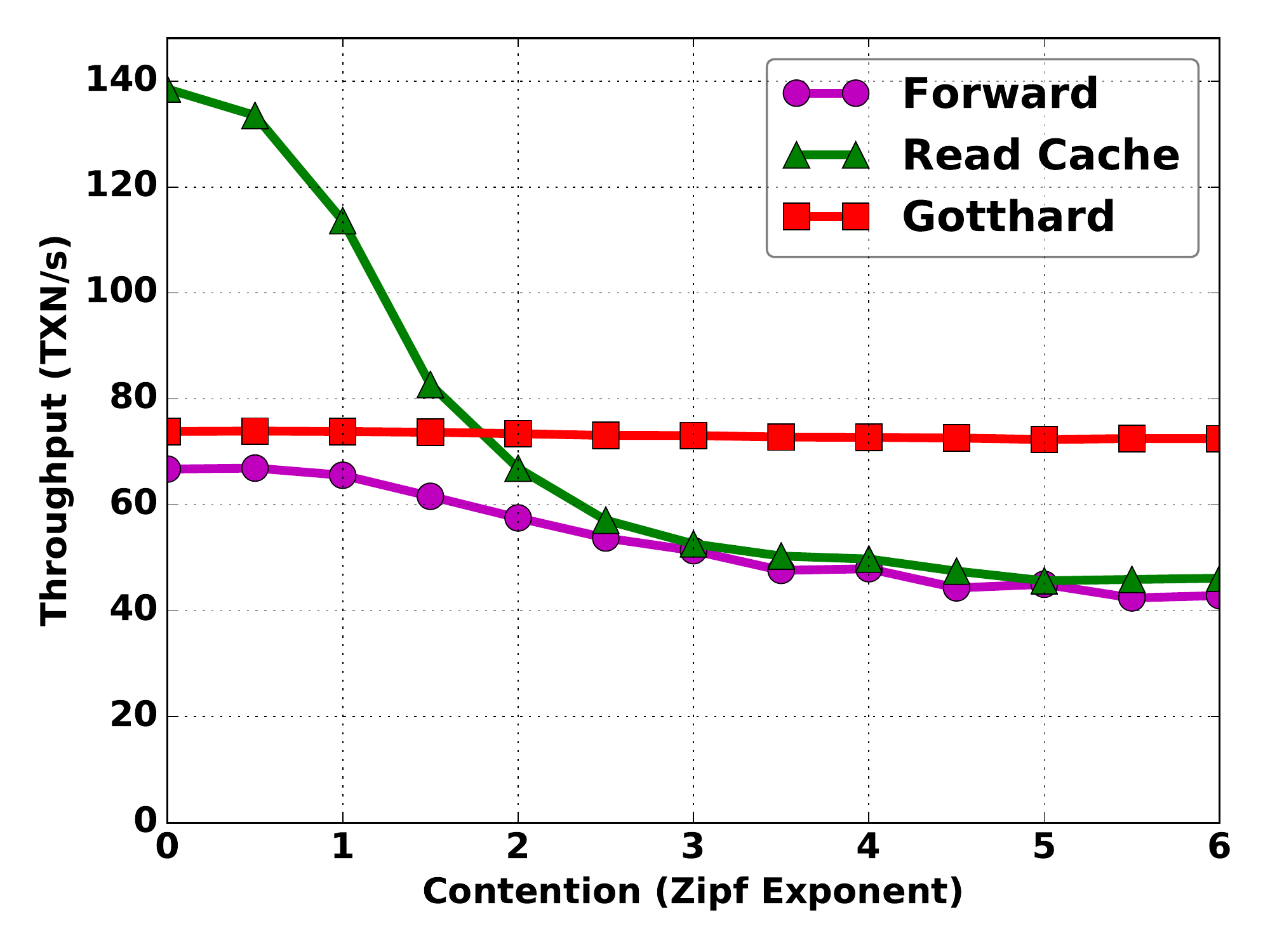}
    \caption{contention vs. throughput}
    \label{fig:rate-vs-zipf}
  \end{subfigure}
  \begin{subfigure}[t]{.45\textwidth}
    \centering
    \includegraphics[width=\textwidth]{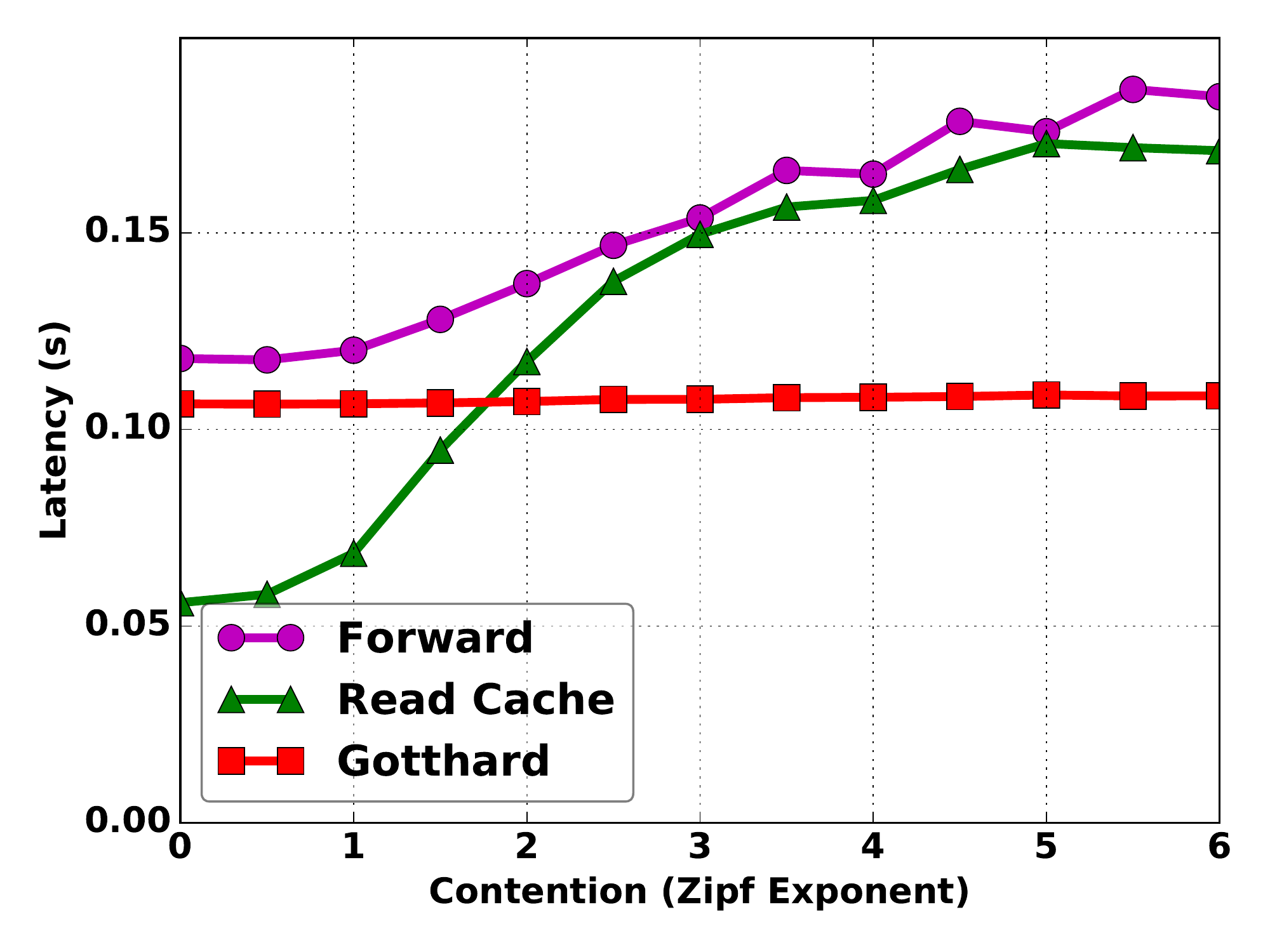}
    \caption{contention vs. latency}
    \label{fig:latency-vs-zipf}
  \end{subfigure}
  \vspace{-1mm}
\caption{Performance with increasing contention.}
\vspace{-2mm}
\end{figure}

\paragraph*{Contention.}
This experiment characterizes the effect of contention on the performance of
Gotthard.  Clients submit transactions that can access one of 10 keys, and the
popularity of each key is dictated by a Zipf distribution.
Figure~\ref{fig:rate-vs-zipf} shows how increasing the Zipf exponent affects
throughput.  With the Zipf exponent at 0, all keys are accessed with the same
probability, and contention increases as the exponent increases.  With low
contention, Read Cache performs comparatively better than Gotthard. However, as
contention increases, the number of aborts grows, limiting the performance of
write transactions for Read Cache.  Gotthard, on the other hand, by
optimistically aborting and retrying transactions closer to the client, is
virtually unaffected by the increase in contention.

\begin{figure}[t]
    \centering
    \begin{minipage}{.45\textwidth}
        \centering
         \includegraphics[width=\textwidth]{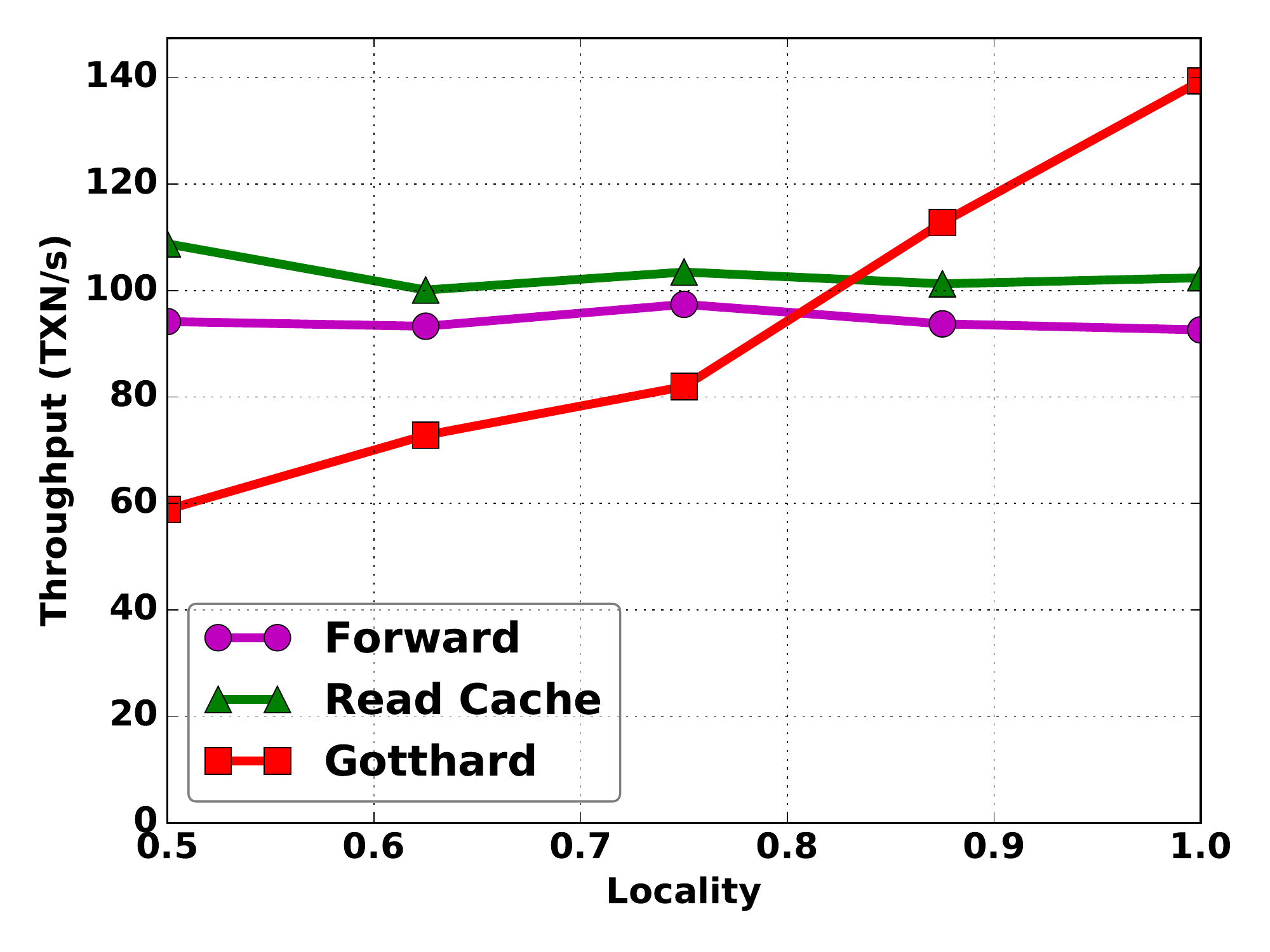}
        \vspace{-1mm}
        \caption{Effect of workload locality on throughput.\\(Multiple switch topology.)}
        \label{fig:locality-experiment}
        \vspace{-2mm}
    \end{minipage}%
    \hfill
    \begin{minipage}{.55\textwidth}
        \centering
        \includegraphics[width=\textwidth]{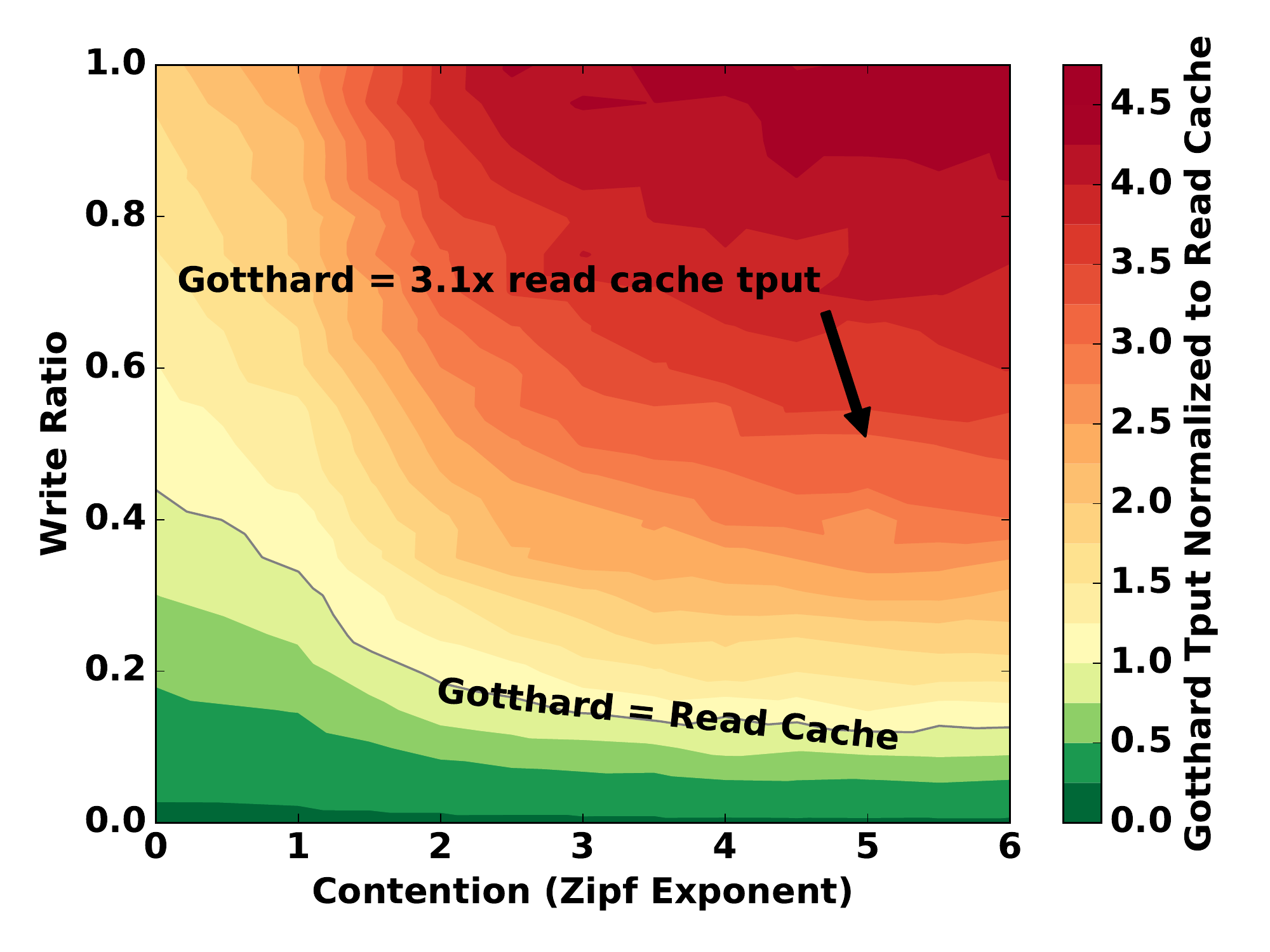}
        \vspace{-1mm}
        \caption{Gotthard's throughput relative to Read Cache as contention and \%writes change. Gotthard does better the closer it is to red (upper right corner).}
          \label{fig:write-vs-zipf}
         \vspace{-2mm}
    \end{minipage}
\end{figure}

In Figure~\ref{fig:write-vs-zipf}, we try to characterize the relative
performance of Gotthard against the Read Cache approach, under varying \%writes
and Zipf exponent. The heatmap shows the throughput of Gotthard normalized to
that of the Read Cache.
Gotthard's relative throughput is higher the closer we get to the upper-right corner of the heatmap.
With low contention (Zipf exponent at 0), Gotthard has better throughput when \%writes is higher than 0.4.
As contention increases, Gotthard has better throughput even when \%writes get as low as 0.15.

\paragraph*{Locality.}
For the locality experiment, we change the topology of our setup to that
illustrated by Figure~\ref{fig:locality}.  In this setup, we have two groups of
8 clients, with each group connected to a different switch.  The third switch,
placed in between the first two switches and the store, is virtually collocated
with the store; there is no added latency between the middle switch and the
store.  We divide a total of 10 keys into two sets of 5 keys, each set being
``local'' to one of the group of clients.  To control locality, we vary the
number of clients on each group that accesses ``remote'' keys, that is, those of
the opposite group.  Thus, when locality is 1, every client accesses keys in the
local set.  At the other extreme, when locality is 0.5, half of the clients in
each group access non-local keys.  Furthermore, keys inside a set are chosen
with probability given by a Zipf distribution of exponent 3.

In Figure~\ref{fig:locality-experiment} we can see that, as expected, locality has a marked effect on Gotthard's performance.
In this scenario, Gotthard does worse than the other two approaches when locality is lower than around 0.8.
Still, as locality increases, so does Gotthard's performance.
Gotthard should be suitable for more sophisticated deployments as long as the expected workload exhibits some degree of locality.

\paragraph*{Summary.}
Overall, the experiments show that Gotthard consistently improves both throughput and average latency in workloads exhibiting contention.
Specifically, \textit{the comparative improvement in performance provided by Gotthard increases as contention goes up.}
In more elaborate setups involving multiple switches, we expect Gotthard to still provide performance benefits, as long as clients exhibit some degree of locality in their access patterns.

\subsection{TPC-C}

TPC-C~\cite{tpcc} is a widely-used benchmark for transaction processing
systems. It models an online transaction processing (OLTP) workload for a
fictional wholesale parts supplier that maintains a number of warehouses for different
sales districts. Clients issue one of five different transactions against the
store to: enter and deliver orders, record payments, check an order status, and
monitor inventory at warehouses.

The TPC-C specification states that the benchmark should be run with a set of
specific parameter values: 1 warehouse, 10 districts, 3000 customers and 100,000
items. For our evaluation, we differed from the standard parameters to use the
following settings: 1 warehouse, 2 districts, 10 customers and 50 items.
We altered these parameters to increase the contention on the data store. 

As a TPC-C driver, we modified an open source Python implementation from Pavlo et al.
that was originally used to issue transactions against 
MongoDB.\footnote{\url{https://github.com/apavlo/py-tpcc}} Since the store
does not have indexes, we modified the transaction executor to only perform
exact selects. To represent the TPC-C database in our store, we map each record
to a key-value pair. The client driver keeps a copy of all records it has read
or written, essentially mirroring the store with a local cache. This eliminates
the unnecessary latency of issuing read requests for records that have not
changed. Consistency is guaranteed by validating the transaction before
committing, ensuring the read values (possibly from the local cache) were
up-to-date.

\paragraph*{Complete benchmark.}

\begin{figure}[t!]
   \centering 
   \begin{subfigure}[t]{.45\textwidth}
     \centering
     \includegraphics[width=\textwidth]{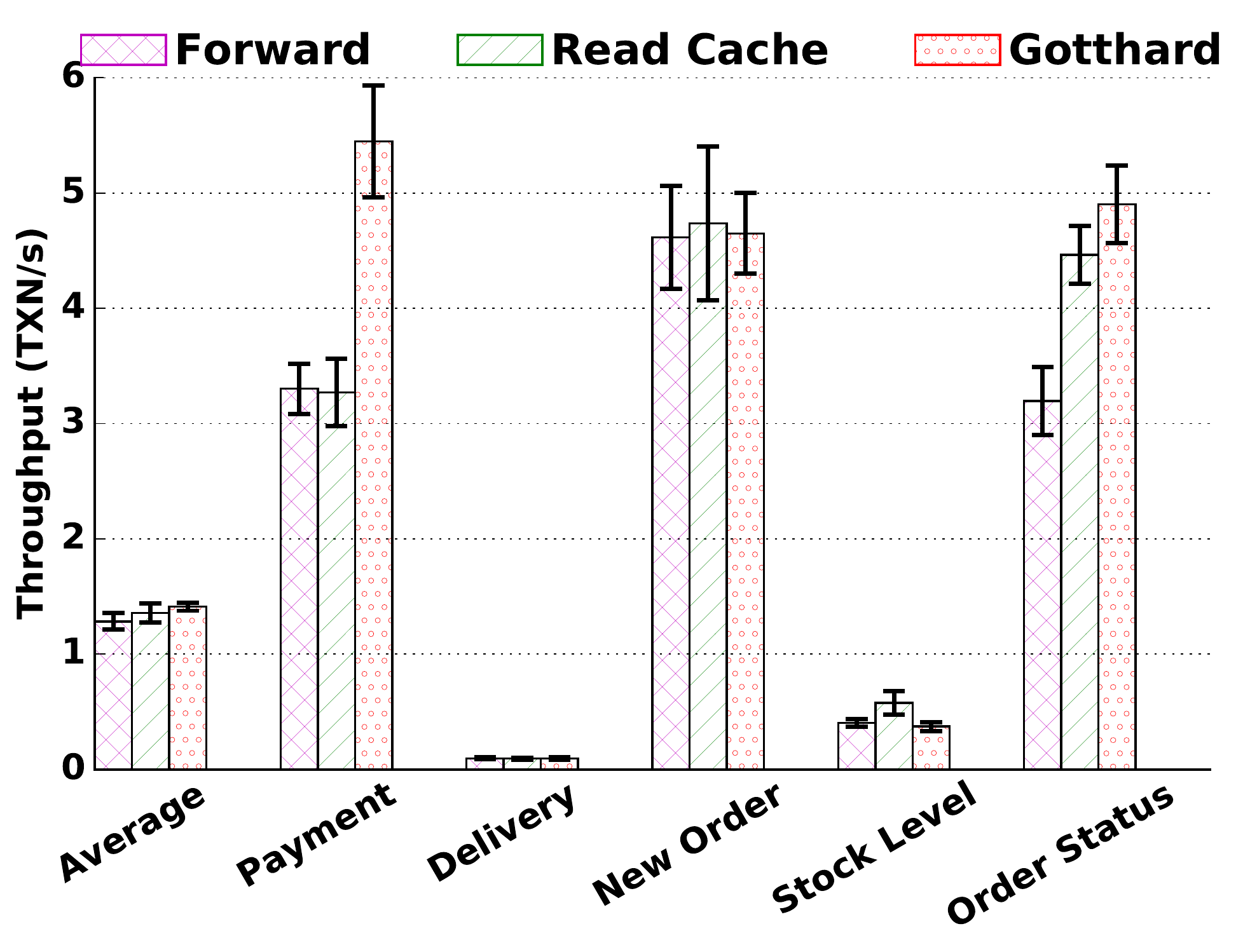}
     \caption{Mininet}
     \label{fig:tpcc}
    \end{subfigure}
   \begin{subfigure}[t]{.45\textwidth}
     \centering
     \includegraphics[width=\textwidth]{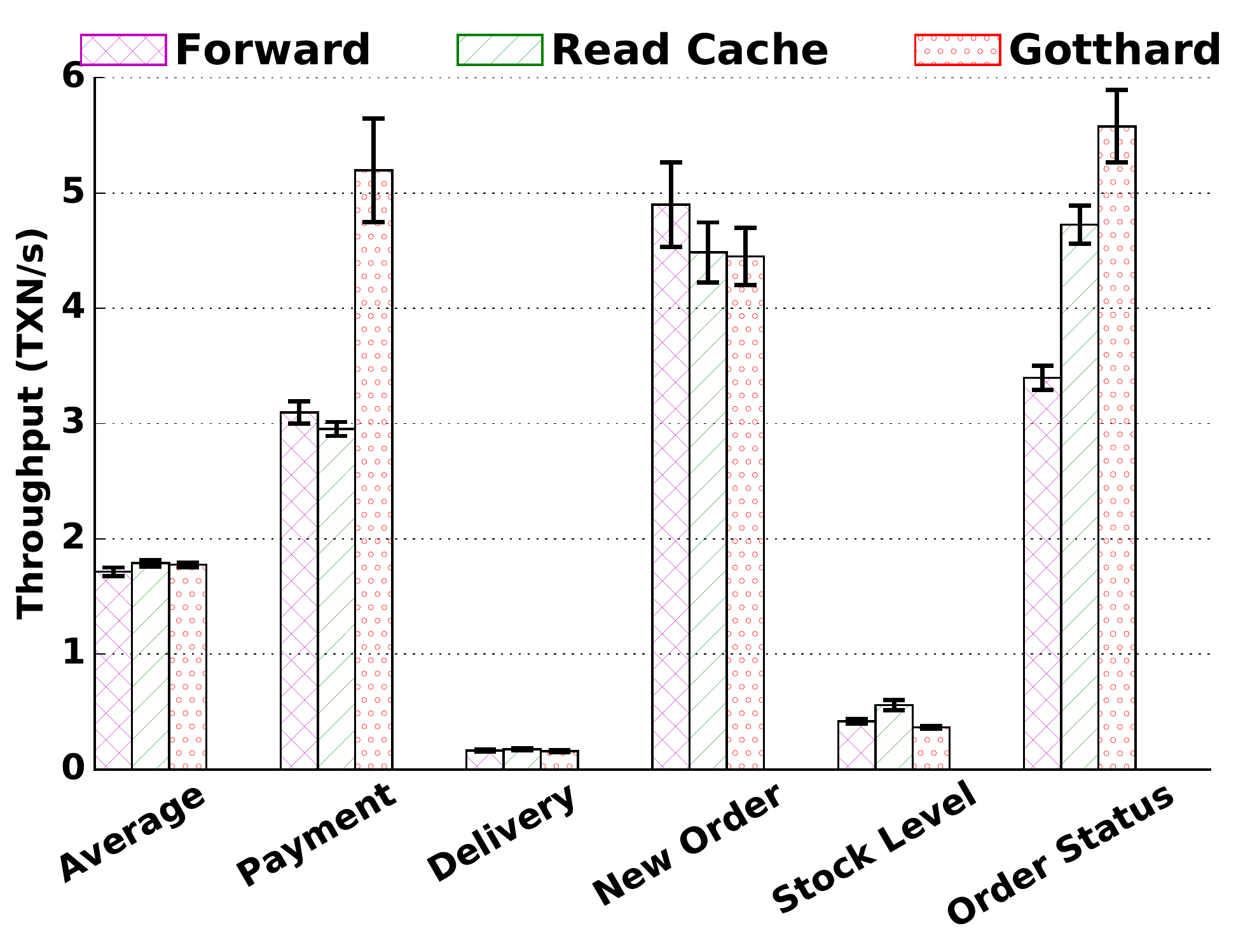}
     \caption{EC2}
     \label{fig:ec2_tpcc}
    \end{subfigure}
\caption{Summary of TPC-C transaction throughput. Gotthard improves the
  throughput of the Payment and Order Status transactions, which contain many write operations.}
\label{fig:tpcc-summary}
\vspace{-1mm}
\end{figure}

Figure~\ref{fig:tpcc-summary} shows a summary of the throughput for the three
switch approaches. The bars on the left show the average of all transactions,
weighted by their frequency. The remaining bars show the throughput for
individual transactions. The results on Mininet (Figure~\ref{fig:tpcc}) and on
EC2 (Figure~\ref{fig:ec2_tpcc}) are similar. Both show that Gotthard improves
throughput for Payment and Order Status. It is less effective for the other
three transactions.

The reason that Gotthard is less effective for Delivery, New Order, and Stock
Level is because when the client receives an abort response for one of these
transactions, it must issue multiple read transactions before they can resubmit.
These read transactions are due to the fact that records in these transactions
have dependencies on records that are accessed by other transactions.

However, we see that on average, Gotthard improves the overall performance. In
the case of the two transactions that can be re-submitted without additional
reads, Gotthard significantly improves the throughput compared to the Read Cache
(around 2x for Payment, and 1.1x for Order status). Below, we focus on these two
transactions in detail.

\begin{figure*}[ht!]
\centering
    \captionsetup{justification=centering}

    \begin{subfigure}[t]{.25\textwidth}
     \centering
     \includegraphics[width=\columnwidth]{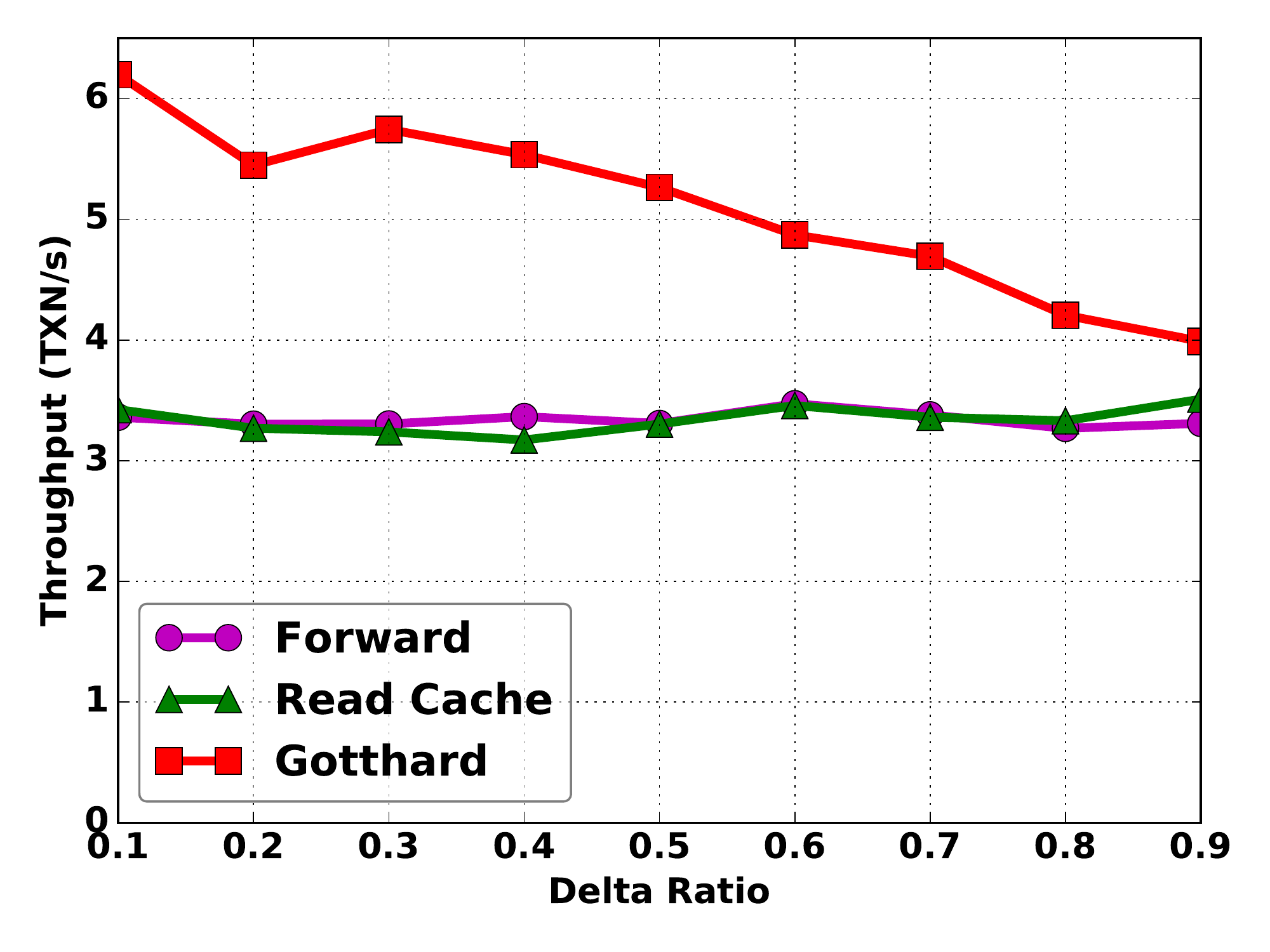}
        \caption{Payment: delta vs. throughput (Mininet)}
     \label{fig:payment-delta-throughput}
   \end{subfigure}%
\begin{subfigure}[t]{.25\textwidth}
     \centering
      \includegraphics[width=\columnwidth]{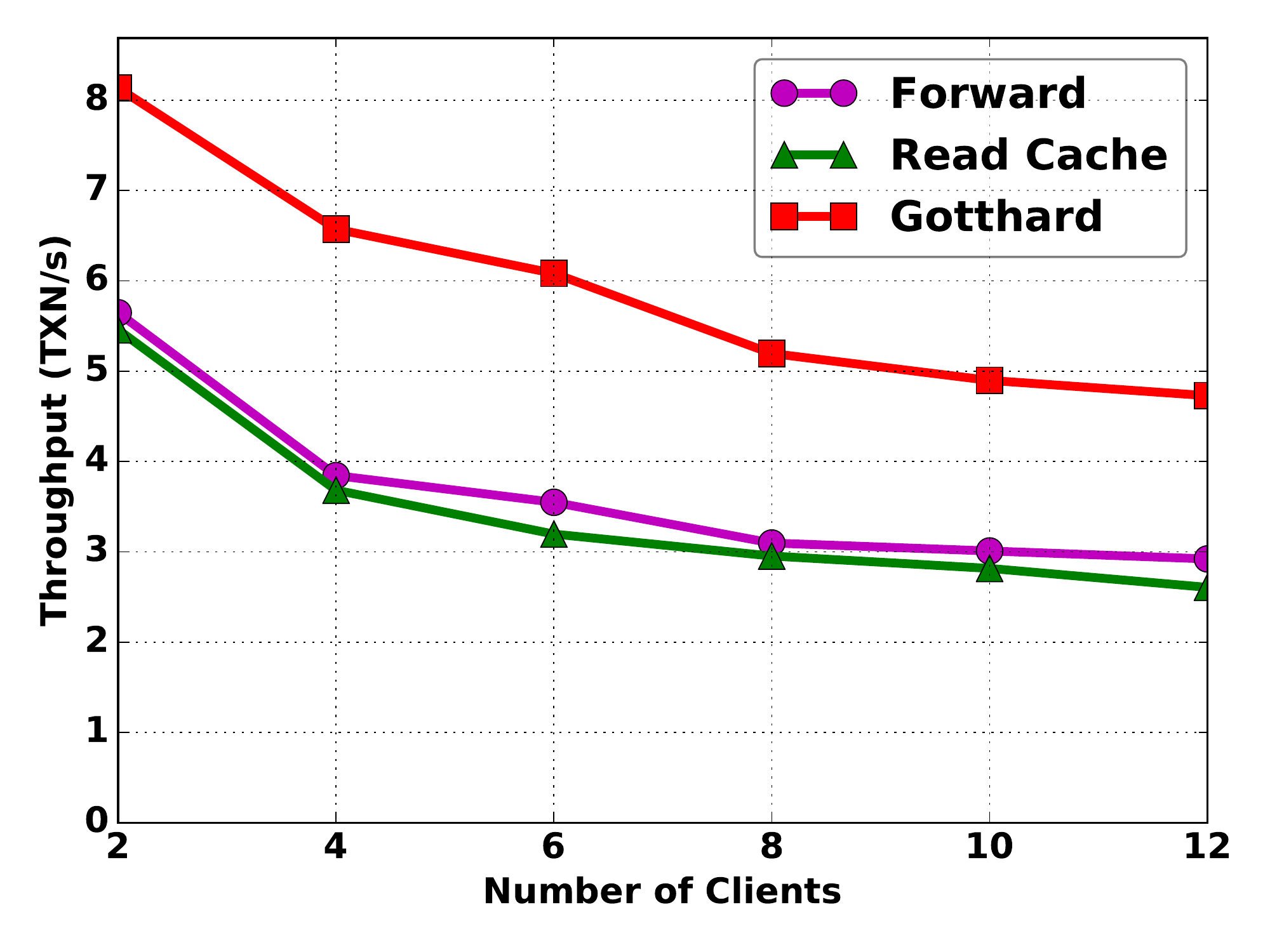}
     \caption{Payment: \#clients vs.\\ throughput (EC2)}
     \label{fig:payment-clients-throughput}
   \end{subfigure}%
   \begin{subfigure}[t]{.25\textwidth}
     \centering
     \includegraphics[width=\columnwidth]{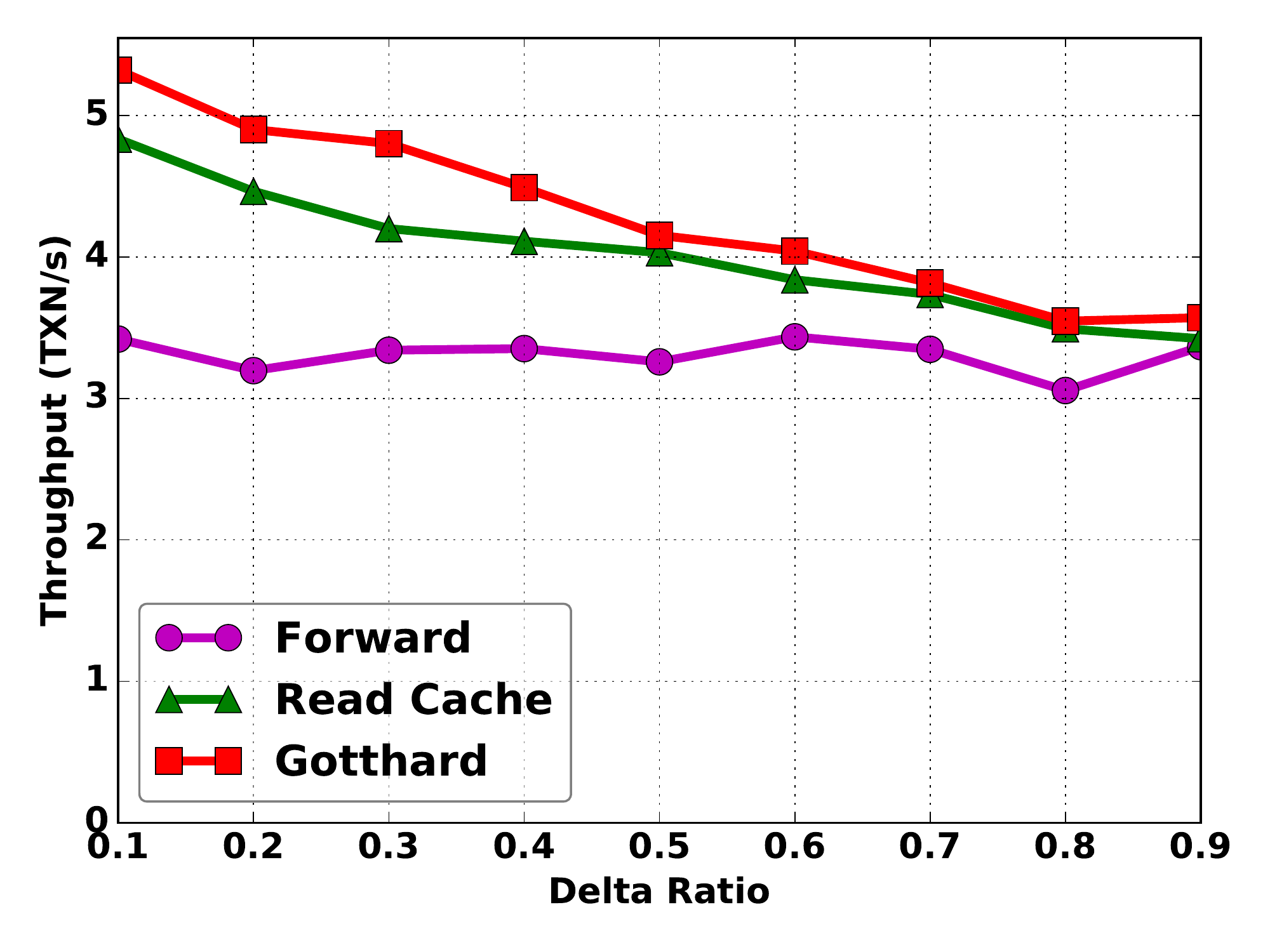}
      \caption{Order Status: delta vs. throughput (Mininet)}
      \label{fig:status-delta-throughput}
    \end{subfigure}%
     \begin{subfigure}[t]{.25\textwidth}
     \centering
      \includegraphics[width=\columnwidth]{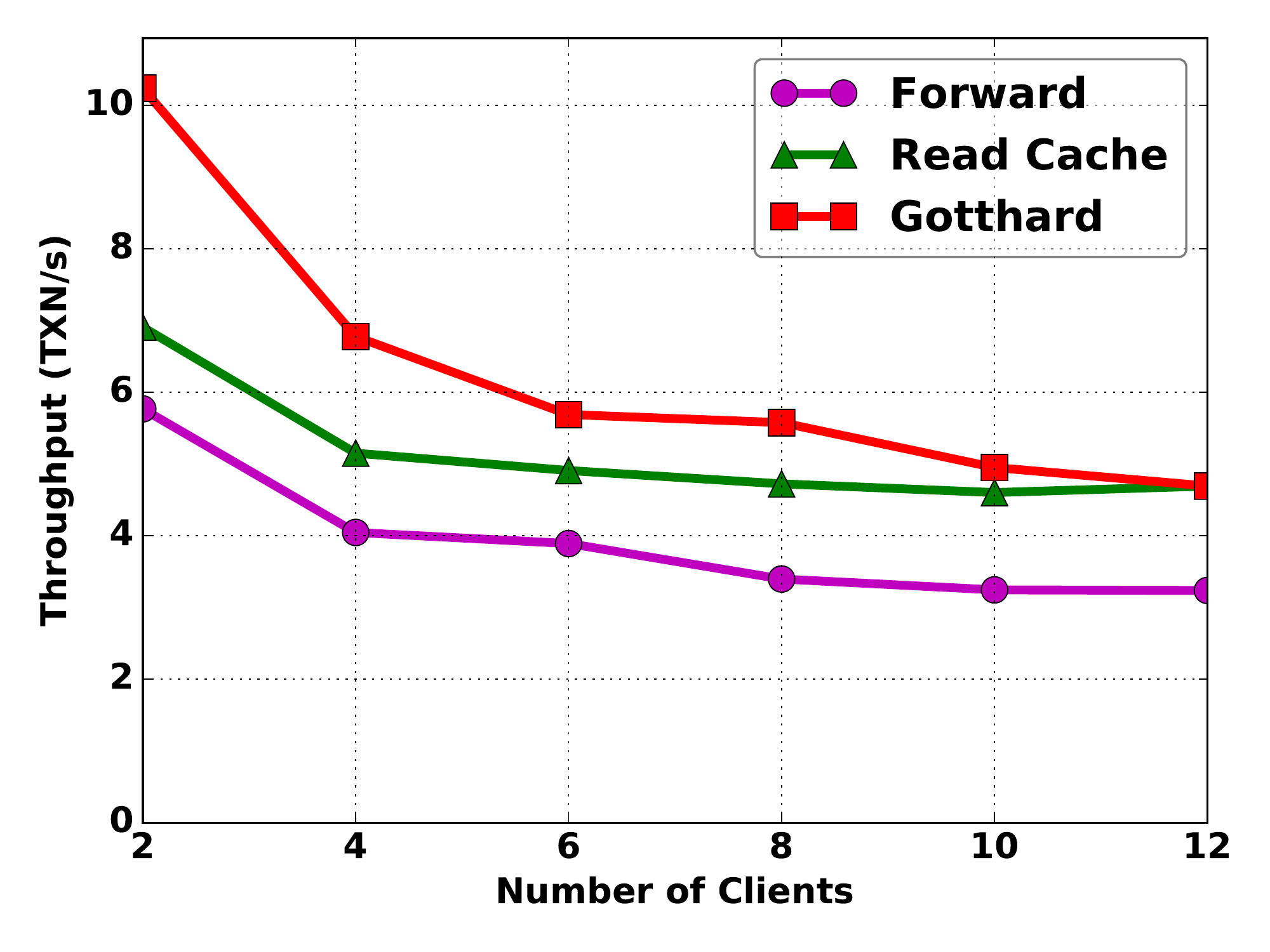}
         \caption{Order Status: \#clients vs. throughput (EC2)}
      \label{fig:status-clients-throughput}
    \end{subfigure}\\
\begin{subfigure}[t]{.25\textwidth}
     \centering
     \includegraphics[width=\columnwidth]{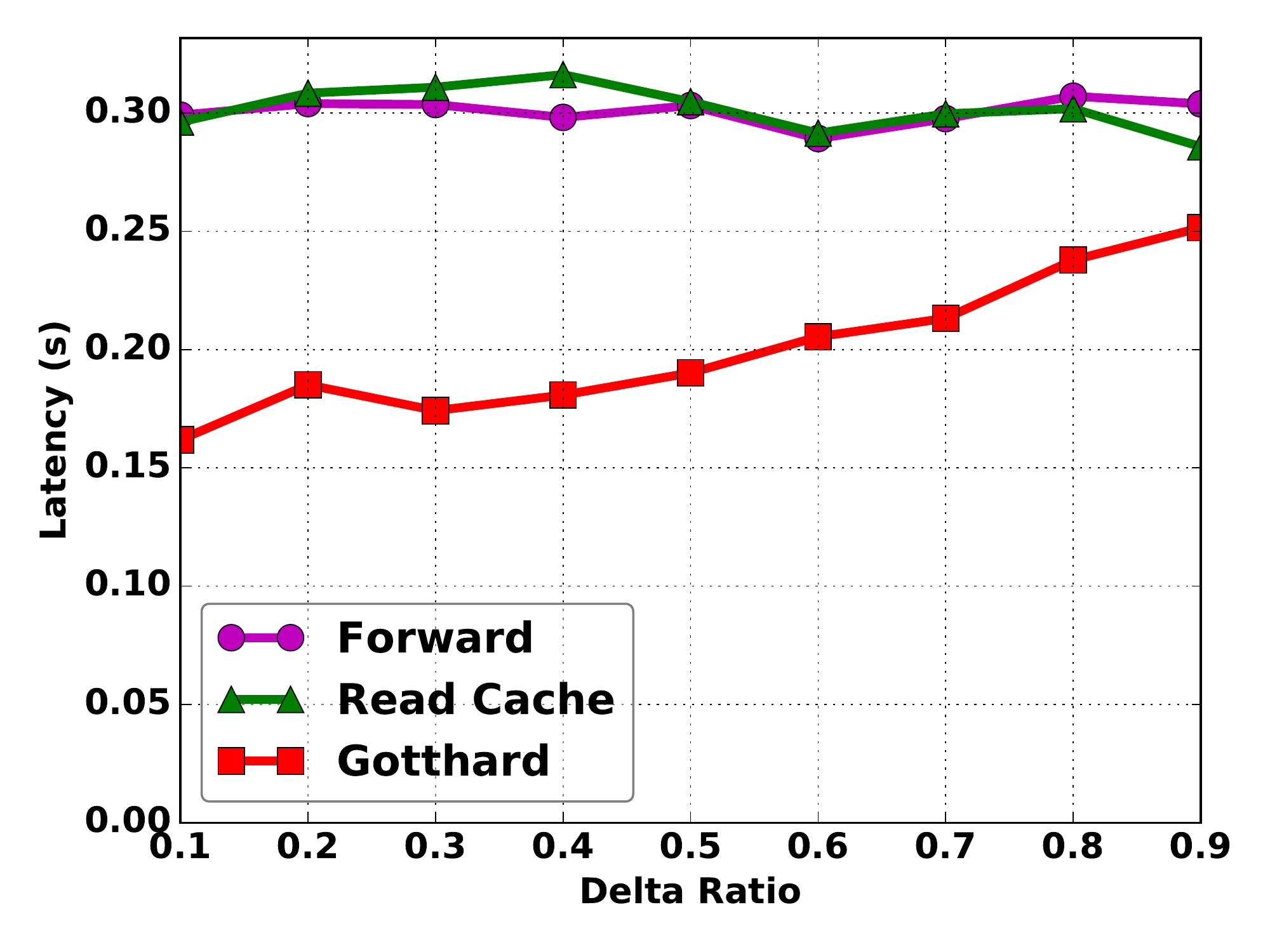}
     \caption{Payment delta: vs. latency (Mininet)}
     \label{fig:payment-delta-latency}
   \end{subfigure}%
   \begin{subfigure}[t]{.25\textwidth}
     \centering
      \includegraphics[width=\columnwidth]{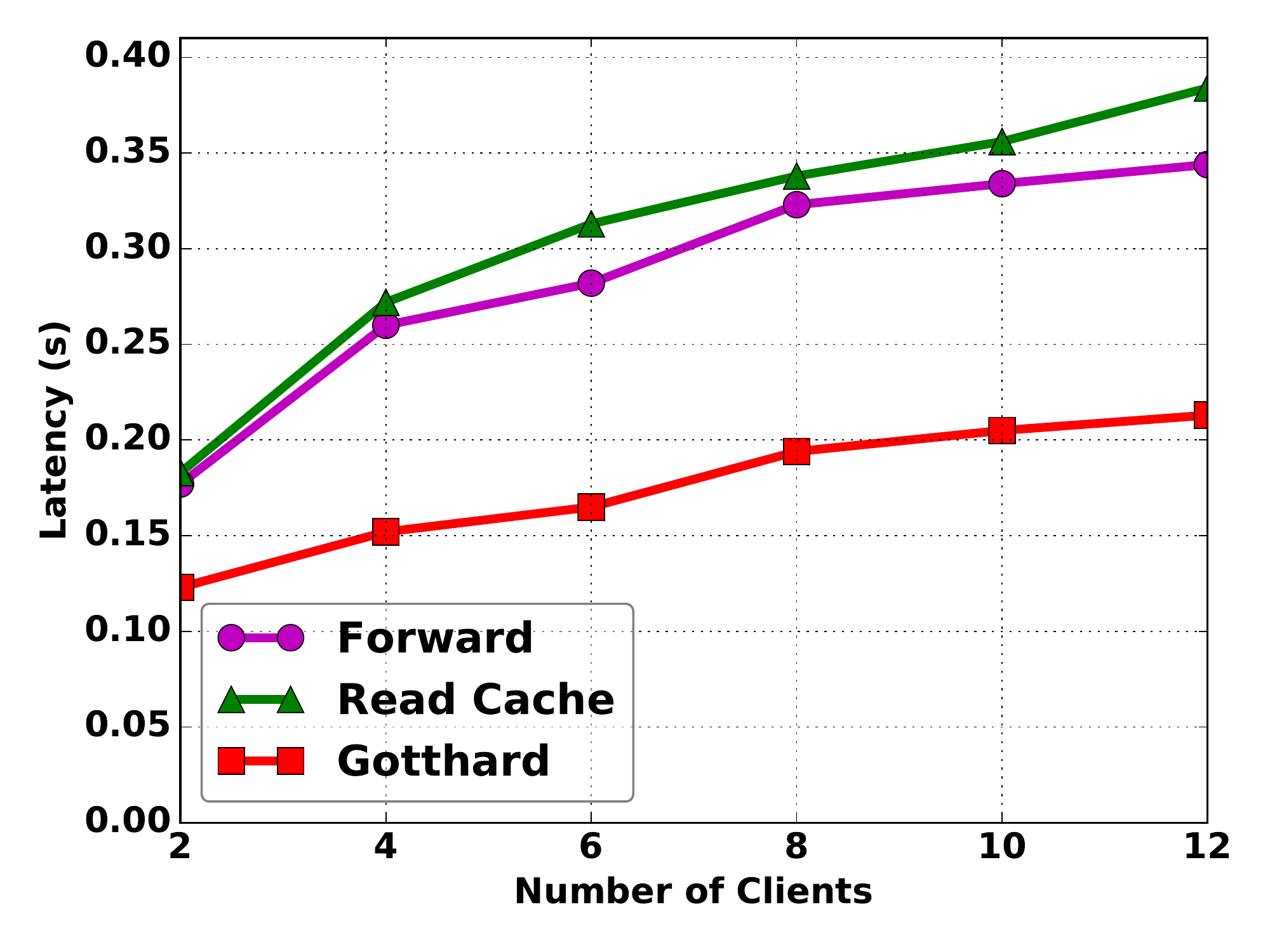}
       \caption{Payment: \#clients vs. latency (EC2)}
     \label{fig:payment-clients-latency}
    \end{subfigure}%
    \begin{subfigure}[t]{.25\textwidth}
     \centering
     \includegraphics[width=\columnwidth]{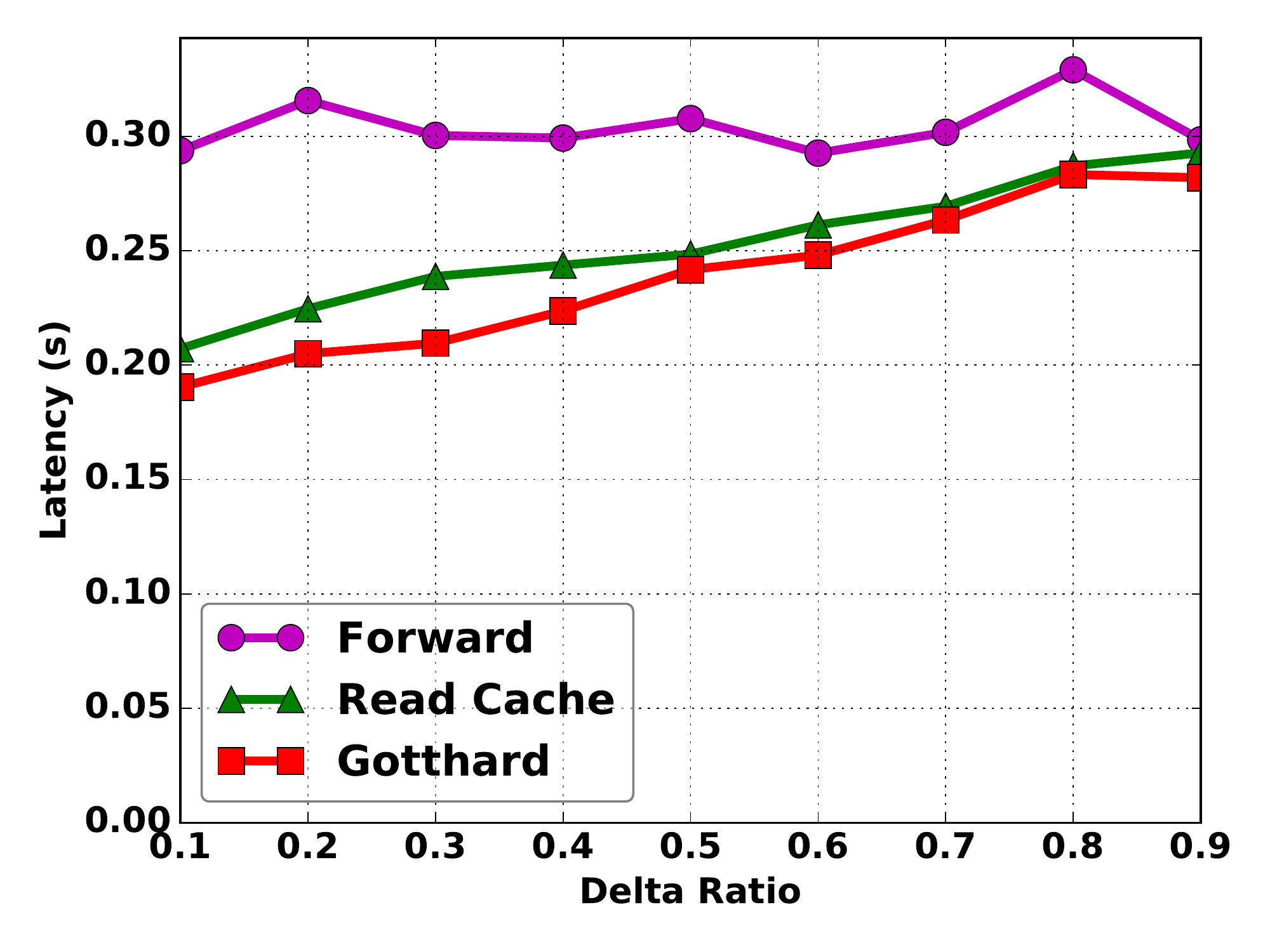}
     \caption{Order Status: delta vs. latency (Mininet)}
     \label{fig:status-delta-latency}
   \end{subfigure}%
   \begin{subfigure}[t]{.25\textwidth}
     \centering
      \includegraphics[width=\columnwidth]{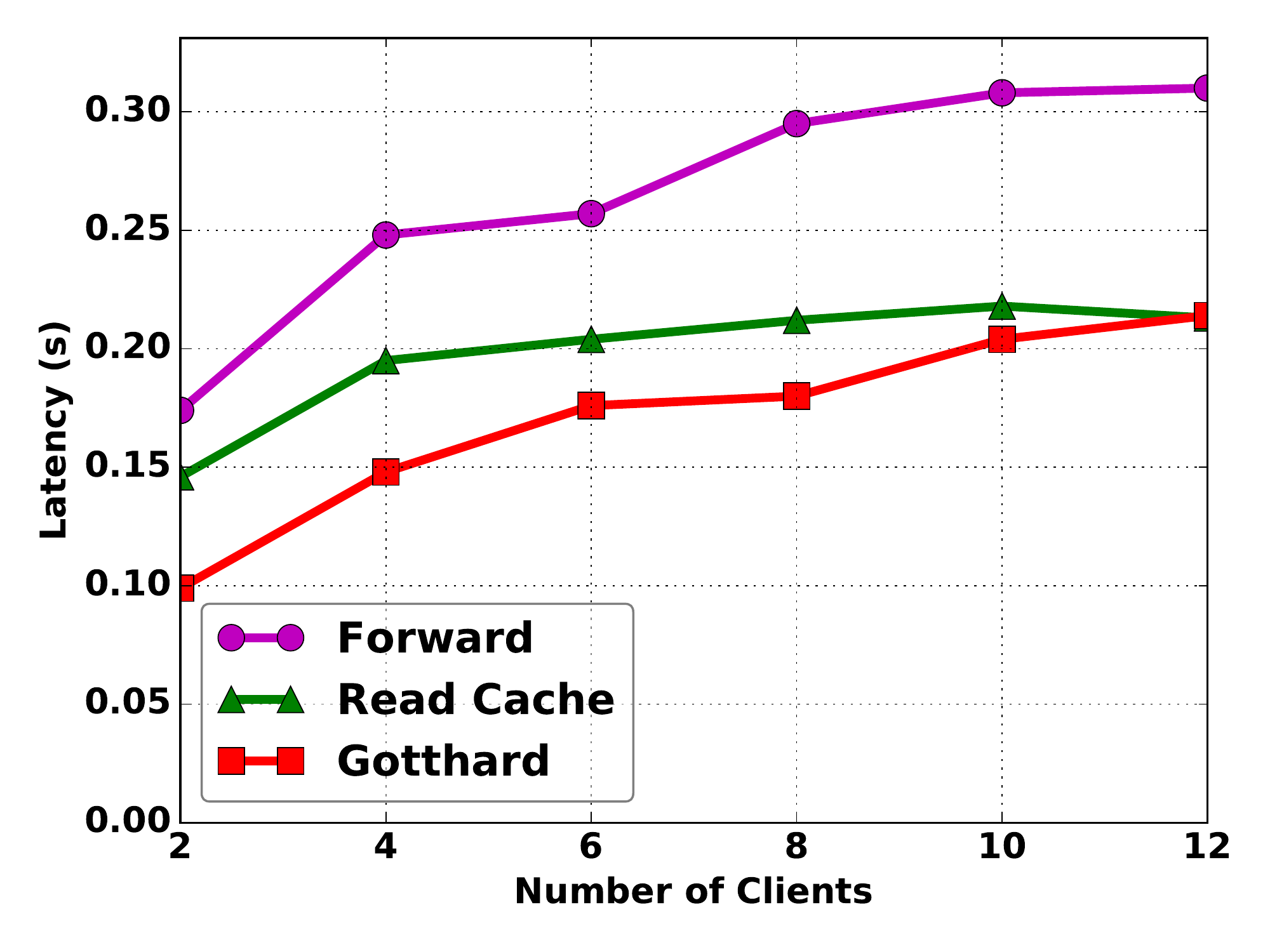}
       \caption{Order Status: \#clients vs. latency (EC2)}
     \label{fig:status-clients-latency}
    \end{subfigure}\\
\caption{TPC-C Payment and Order Status transactions}
\end{figure*}

\paragraph*{Payment transaction.}
Figures~\ref{fig:payment-delta-throughput} and \ref{fig:payment-delta-latency}
show the throughput and latency for the Payment transaction as we vary the
delta ratio. When the switch is closer to the clients (lower delta ratio),
Gotthard has lower latency and higher throughput than both the forwarding and
Read Cache switches. As the switch is moved further from the clients (higher
delta ratio), Gotthard's performance is more similar to the other switches'.

Figures~\ref{fig:payment-clients-throughput} and \ref{fig:payment-clients-latency}
show the throughput and latency for the Payment transaction as we vary the
number of clients. Gotthard has higher throughput and lower latency as the
number of clients increases. The other switches saturate at 8 clients, while
Gotthard continues to scale.

\paragraph*{Order Status transaction.}
Figures~\ref{fig:status-delta-throughput} and \ref{fig:status-delta-latency}
show the throughput and latency for the Order Status transaction as we vary the
delta ratio. Gotthard has higher throughput than the Read Cache until the switch
is halfway between the clients and the store (delta ratio 0.5). Gotthard
always has lower latency than the others.

Figures~\ref{fig:status-clients-throughput} and \ref{fig:status-clients-latency}
show the throughput and latency as the number of clients is increased. The other
switches saturate at 4 clients, while Gotthard maintains higher throughput until
10 clients.

\paragraph*{TPC-C Summary.}
Overall, the experiments show that Gotthard has higher throughput than the Read
Cache for high contention transactions. Gotthard has the greatest benefit for
the TPC-C workload when the switch is closer to the client, and still provides
better performance when the switch is close to the store.  As the number of
clients increases, and thus the contention, Gotthard reaches a higher saturation
point after the other switches.

\section{Related Work}
\label{sec:related}

\paragraph*{Proxies and caches}
The idea of using a proxy to extend distributed services is a well-established
idea~\cite{shapiro86} that has been widely
adopted~\cite{brewer98,anderson96,joseph95, knutsson03}.  Proxies are often used
to scale services by caching copies of data closer to clients, such as with
content distribution networks (CDNs)~\cite{nottingham01,freedman04,wang02}. CDNs
typically are used for static content, although there are examples of proxies
used for dynamic content~\cite{grimm06}.

Prior work has also explored the possibility of leveraging the network to route
requests dynamically to proxies to service requests~\cite{wang02}. Notably,
recent work on SwitchKV~\cite{li16} uses OpenFlow-enable switches to dynamically
route read requests to proxy caches.

Gotthard differs from this work in that it is not a cache, \emph{per se}. It
keeps copies of transaction requests, but it does not service client read
requests. Rather, it uses copies of previous requests to make informed decisions
about when to abort transactions early, with the goal of reducing latency for
write-heavy workloads.

\paragraph*{Geo-distributed databases.}
Many recent works deal with geo-distribution in the context of transactional
databases, most with a focus on replication and partitioning.  Works such as
\cite{baker11megastore, corbett12, kraska2013mdcc, nawab2015minimizing,
sciascia13} aim at providing strong consistency (i.e., serializability) over
wide-area networks.  To improve latency and availability, many works also
propose weaker consistency criteria such as Parallel Snapshot
Isolation~\cite{sovran2011transactional}, causal consistency~\cite{lloyd2011don,
lloyd2013stronger} and RAMP transactions~\cite{bailis2014scalable}.
PLANET~\cite{pang14} exposes transaction state to the application, enabling
speculative processing and faster revocation using the \emph{guess and
apologies} paradigm~\cite{helland2009building}.  Gotthard's approach is
complimentary to the aforementioned research, and combining our approach with a
full-fledged database solution is part of our future work.

\paragraph*{Data plane programming languages.}
Gotthard is written in P4~\cite{bosshart14}, although there are several other
 projects have proposed domain-specific languages for data plane
 programming. Notable examples including Huawei's POF~\cite{song13} and Xilinx's
PX~\cite{brebner14}. We chose to focus on P4 because there is a growing
 community of active users, and it is relatively more mature than the other
 choices. However, the ideas for implementing Gotthard should
 generalize to other languages.

\paragraph*{Application logic in the network.}
Several recent projects investigate leveraging network programmability for
improved application performance.  One thread of research has focused on
improving application performance through traffic management. Examples of such
systems include PANE~\cite{ferguson13}, EyeQ~\cite{jeyakumar13a}, and
Merlin~\cite{soule14} which all use resource scheduling to improve job
performance. NetAgg~\cite{mai14} leverages user-defined \emph{combiner}
functions to reduce network congestion.  Another thread of research has focused
on moving application logic into network devices.  Dang et al.~\cite{dang15}
proposed the idea of moving consensus logic in to network devices.  Paxos Made
Switch-y~\cite{dang16} describes an implementation of Paxos in P4. Istv\'{a}n et
al.~\cite{istvan16} implement Zookeeper's atomic broadcast
 on an FPGA.

\section{Conclusion}
\label{sec:conclusion}

The advent of flexible hardware and expressive data\-plane programming languages
will have a profound impact on networks. One possible use of this emerging
technology is to move logic traditionally associated with the application layer
into the network itself.

With Gotthard, we have leveraged this technology to move transaction processing
logic into network devices. Gotthard uses a custom packet header and
programmable switches to identify doomed transactions, and abort them in the
network as early as possible.

For write-intensive, high-contention workloads, significantly reduce the overall
latency for processing transactions. Furthermore, Gotthard reduces load on the
store, increasing system throughput.

Overall, Gotthard compliments prior techniques for reducing network latency,
such as using a cache to service read requests. Moreover, Gotthard provides a
novel application of data plane programming languages that advances the
state-of-the-art in this emerging area of research.

{\small
\bibliographystyle{abbrv}
\balance
\bibliography{main}
}

\end{document}